\documentclass[twocolumn]{aastex63}
\pdfoutput=1
\usepackage{array}
\usepackage{amsmath}
\usepackage[counterclockwise, figuresleft]{rotating}
\usepackage{txfonts}
\usepackage{booktabs}
\usepackage{multirow}
\usepackage[caption=false]{subfig}

\newcommand{\mearth}{M_\oplus}
\newcommand{\rearth}{R_{\rm \oplus}}
\newcommand{\msun}{M_\odot}
\newcommand{\rsun}{R_{\rm \odot}}
\newcommand{\lsun}{L_\odot}
\newcommand{\rhosun}{\rho_\odot}

\newcommand{\teff}{T_\mathrm{eff}}
\newcommand{\logg}{\log g}

\def\m2s2{\hbox{\,m$^{2}$\,s$^{-2}$}} 
\def\kms{\hbox{\,km\,s$^{-1}$}}       
\def\vsini{\hbox{$v$\,sin\,$i$}}      
\def\sini{\hbox{sin\,$i$}}      

\begin{document}


\title{K2-79b and K2-222b: Mass measurements of two small exoplanets with periods beyond 10 days that overlap with periodic magnetic activity signals}


\author[0000-0001-8838-3883]{Chantanelle Nava}
\author[0000-0003-3204-8183]{Mercedes L\'{o}pez-Morales}
\affiliation{Center for Astrophysics ${\rm \mid}$ Harvard {\rm \&} Smithsonian, 60 Garden Street, Cambridge, MA 02138, USA}

\author[0000-0001-7254-4363]{Annelies Mortier}
\affiliation{Astrophysics Group, Cavendish Laboratory, University of Cambridge, J.J. Thomson Avenue, Cambridge CB3 0HE, UK}
\affiliation{Kavli Institute for Cosmology, University of Cambridge, Madingley Road, Cambridge CB3 0HA, UK}

\author[0000-0003-1957-6635]{Li Zeng}
\affiliation{Center for Astrophysics ${\rm \mid}$ Harvard {\rm \&} Smithsonian, 60 Garden Street, Cambridge, MA 02138, USA}
\affiliation{Department of Earth and Planetary Sciences, Harvard University, 20 Oxford Street, Cambridge, MA 02138, USA}

\author[0000-0001-6777-4797]{Helen A.C. Giles}
\affiliation{Observatoire de Gen\`eve, Universit\'e de Gen\`eve, 51 ch. des Maillettes, CH-1290 Sauverny, Switzerland}

\author[0000-0001-6637-5401]{Allyson Bieryla}
\affiliation{Center for Astrophysics ${\rm \mid}$ Harvard {\rm \&} Smithsonian, 60 Garden Street, Cambridge, MA 02138, USA}

\author[0000-0001-7246-5438]{Andrew Vanderburg}
\affiliation{Department of Astronomy, University of Wisconsin - Madison, WI 53706, USA}

\author[0000-0003-1605-5666]{Lars A. Buchhave}
\affiliation{DTU Space, National Space Institute, Technical University of Denmark, Elektrovej 328, DK-2800 Kgs. Lyngby, Denmark}

\author[0000-0003-1200-0473]{Ennio Poretti}
\affiliation{Fundaci\'on Galileo Galilei-INAF, Rambla J. A. F. Perez, 7, E-38712, S.C. Tenerife, Spain}
\affiliation{INAF-Osservatorio Astronomico di Brera, via E. Bianchi 46, 23807 Merate (LC), Italy}

\author[0000-0001-7032-8480]{Steven H. Saar}
\affiliation{Center for Astrophysics ${\rm \mid}$ Harvard {\rm \&} Smithsonian, 60 Garden Street, Cambridge, MA 02138, USA}

\author[0000-0002-9332-2011]{Xavier Dumusque}
\affiliation{Observatoire de Gen\`eve, Universit\'e de Gen\`eve, 51 ch. des Maillettes, CH-1290 Sauverny, Switzerland}

\author[0000-0001-9911-7388]{David W. Latham}
\affiliation{Center for Astrophysics ${\rm \mid}$ Harvard {\rm \&} Smithsonian, 60 Garden Street, Cambridge, MA 02138, USA}

\author[0000-0002-9003-484X]{David Charbonneau}
\affiliation{Center for Astrophysics ${\rm \mid}$ Harvard {\rm \&} Smithsonian, 60 Garden Street, Cambridge, MA 02138, USA}

\author[0000-0001-9984-4278]{Mario Damasso}
\affiliation{INAF - Osservatorio Astrofisico di Torino, Via Osservaorio 20, I10025 Pino Torinese, Italy}

\author[0000-0002-6177-198X]{Aldo S. Bonomo}
\affiliation{INAF - Osservatorio Astrofisico di Torino, Via Osservaorio 20, I10025 Pino Torinese, Italy}

\author[00000000-0001-7120-5837]{Christophe Lovis}
\affiliation{Observatoire de Gen\`eve, Universit\'e de Gen\`eve, 51 ch. des Maillettes, CH-1290 Sauverny, Switzerland}

\author[0000-0002-8863-7828]{Andrew Collier Cameron}
\affiliation{Centre for Exoplanet Science, SUPA, School of Physics and Astronomy, University of St Andrews, St Andrews KY16 9SS, UK}

\author[0000-0003-3773-5142]{Jason D. Eastman}
\affiliation{Center for Astrophysics ${\rm \mid}$ Harvard {\rm \&} Smithsonian, 60 Garden Street, Cambridge, MA 02138, USA}

\author[0000-0002-7504-365X]{Alessandro Sozzetti}
\affiliation{INAF - Osservatorio Astrofisico di Torino, Via Osservaorio 20, I10025 Pino Torinese, Italy}

\author[0000-0003-1784-1431]{Rosario Cosentino}
\affiliation{Fundaci\'on Galileo Galilei-INAF, Rambla J. A. F. Perez, 7, E-38712, S.C. Tenerife, Spain}

\author[0000-0000-0000-0000]{Marco Pedani}
\affiliation{Fundaci\'on Galileo Galilei-INAF, Rambla J. A. F. Perez, 7, E-38712, S.C. Tenerife, Spain}

\author[0000-0002-5254-6289]{Francesco Pepe}
\affiliation{Observatoire de Gen\`eve, Universit\'e de Gen\`eve, 51 ch. des Maillettes, CH-1290 Sauverny, Switzerland}

\author[0000-0002-1742-7735]{Emilio Molinari}
\affiliation{INAF - Osservatorio Astronomico di Cagliari, via della Scienza 5, 09047, Selargius, Italy}

\author[0000-0001-7014-1771]{Dimitar Sasselov}
\affiliation{Center for Astrophysics ${\rm \mid}$ Harvard {\rm \&} Smithsonian, 60 Garden Street, Cambridge, MA 02138, USA}

\author[0000-0002-9352-5935]{Michel Mayor}
\author[0000-0003-0996-6402]{Manu Stalport}
\affiliation{Observatoire de Gen\`eve, Universit\'e de Gen\`eve, 51 ch. des Maillettes, CH-1290 Sauverny, Switzerland}

\author[0000-0002-6492-2085]{Luca Malavolta}
\affiliation{INAF – Osservatorio Astronomico di Padova, Vicolo del l'Osservatorio 5, I-35122 Padova, Italy}
\affiliation{Dipartimento di Fisica e Astronomia ‘Galileo Galilei’, Universit\'a di Padova, Vicolo del l’Osservatorio 3, I-35122 Padova, Italy}

\author[0000-0002-6379-9185]{Ken Rice}
\affiliation{SUPA, Institute for Astronomy, University of Edinburgh, Royal Observatory, Blackford Hill, Edinburgh EH93HJ, UK}
\affiliation{Centre for Exoplanet Science, University of Edinburgh, Edingburgh EH93FD, UK}

\author[0000-0002-9718-3266]{Christopher A. Watson}
\affiliation{Astrophysics Research Centre, School of Mathematics and Physics, Queen’s University Belfast, Belfast, BT7 1NN, UK}

\author[0000-0002-4272-4272]{A. F. Martinez Fiorenzano}
\affiliation{Fundaci\'on Galileo Galilei-INAF, Rambla J. A. F. Perez, 7, E-38712, S.C. Tenerife, Spain}

\author[0000-0000-0000-0000]{Luca Di Fabrizio}
\affiliation{Fundaci\'on Galileo Galilei-INAF, Rambla J. A. F. Perez, 7, E-38712, S.C. Tenerife, Spain}




\begin{abstract}
We present mass and radius measurements of K2-79b and K2-222b, two transiting exoplanets orbiting active G-type stars. Their respective 10.99d and 15.39d orbital periods fall near periods of signals induced by stellar magnetic activity. The two signals might therefore interfere and lead to an inaccurate estimate of exoplanet mass. We present a method to mitigate these effects when radial velocity and activity indicator observations are available over multiple observing seasons and the orbital period of the exoplanet is known. We perform correlation and periodogram analyses on sub-sets composed of each target's two observing seasons, in addition to the full data sets. For both targets, these analyses reveal an optimal season with little to no interference at the orbital period of the known exoplanet. We make a confident mass detection of each exoplanet by confirming agreement between fits to the full radial velocity set and the optimal season. For K2-79b, we measure a mass of 11.8 $\pm$ 3.6 ${\rm M_{\earth}}$ and a radius of 4.09 $\pm$ 0.17 ${\rm R_{\earth}}$. For K2-222b, we measure a mass of 8.0 $\pm$ 1.8 ${\rm M_{\earth}}$ and a radius of 2.35 $\pm$ 0.08 ${\rm R_{\earth}}$. According to model predictions, K2-79b is a highly irradiated Uranus-analog and K2-222b hosts significant amounts of water ice. We also present an RV solution for a candidate second companion orbiting K2-222 at 147.5d.
\vspace{.4in}
\end{abstract}

\section{Introduction}\label{Introduction}
The HARPS radial velocity (RV) survey first revealed the existence of a sub-population of super-Earth and Neptune-like planets in tight orbits \citep{Mayor2008, Lovis2009}.  A few years later, NASA's {\it Kepler} mission found that these small planets, with sizes between that of Earth and Neptune, are the most abundant type of exoplanets with periods less than $\approx$100d \citep[e.g.][]{borucki2011, batalha2013, fressin13}. These types of planets are not found in the Solar System, and their existence was not predicted by contemporary planet-formation models \citep{ida&lin1, ida&lin2, mordasini1, mordasini2}. Subsequent studies, using a greater number of small exoplanets from {\it Kepler}, revealed relatively few planets with radii between 1.5$R_{\Earth}$ and 2$R_{\Earth}$, now commonly referred to as the {\it radius valley} \citep{fulton17, zeng2017a, zeng2017b, vaneylen18}. There also appears to be a scarcity of exoplanets in the {\it sub-Saturnian desert}, with radii larger than 4$R_{\Earth}$ but smaller than a gas giant \citep{zeng18}. Exoplanets in the sub-Saturnian desert appear separated from Sub-Neptunes by the {\it radius cliff}, a steep drop-off in exoplanet occurrence starting near 3$R_{\Earth}$ and attributed to the non-linear solubility of hydrogen in magma at relatively high pressures \citep{Kite2019}.

The exoplanet community has increased efforts to measure masses of small exoplanets to develop a more complete picture of the processes involved in their formation. Bulk densities measured for exoplanets on both sides of the radius valley suggest that those above the gap have very different compositions from those below \citep{rogers15, zeng18, zeng19}. Exoplanets below the gap have relatively high densities, consistent with rocky compositions. Exoplanets above the gap have lower densities that are degenerate, consistent with rocky cores surrounded by envelopes of liquid and/or gas. 

Several mechanisms have been proposed in the literature to explain the observed small planet radius valley: Photo-evaporation, internal heat-powered mass loss, and water-rich vs. water-poor formation. Photo-evaporation refers to the shedding of a planet's envelope from prolonged exposure to its host star \citep{hansen&murray12, mordasini3, mordasini4, alibert13, chiang&laughlin13, chatterjee&tan14, coleman&nelson14, lee14, raymond&cossou14, lee&chiang16}. Like photo-evaporation, core-powered mass-loss also causes an exoplanet to shed its envelope, but is instead powered by radiation from the hot rocky core as it cools and winds in the hot gaseous envelope \citep{owen&wu16, ginzburg16}. Successful models of photo-evaporation, core-powered mass-loss, and atmospheric ``boil off" thus far have assumed that all planets above the gap host Hydrogen/Helium (H or H/He) envelopes, excluding water from their simulations. However, simulations including water and ices have been carried out to form water-rich exoplanets with rocky or rocky/icy cores that accrete significant amounts of water and gaseous envelopes if massive enough \citep{zeng18,zeng19}.



Comparing small planets of varying surface gravity across a range of stellar irradiation levels can help identify driving factors behind the observed population of small exoplanets \citep{lopez18, cloutier20}. So far, the majority of small planets with mass measurements occur at relatively short orbital periods ($P_{\rm  orb}$ $<$ 10 days), and therefore are subject to high irradiation levels. Mass and radius measurements of small planets at longer periods are therefore essential for developing a more complete understanding of small exoplanet formation. 
 
Measuring small exoplanet masses at longer orbital periods presents challenges, including hard-to-detect small-amplitude Doppler signals, and potential interference between periodic signals from stellar activity and $P_{\rm orb}$. The second problem is even more challenging than initially recognized because ${\rm P_{orb}}$ need not overlap perfectly with ${ P_{\rm rot}}$ or one of its multiples/harmonics for the stellar activity and Doppler signals to interfere \citep{Nava2020}. \citet{Mortier2016} tackled the feat of determining the mass on an exoplanet with ${\rm P_{orb}}$ near $P_{\rm rot}$, but generally these cases have been set aside to measure masses of exoplanets with less challenging interference from stellar magnetic active regions. 

In this paper we present a method to mitigate these potential interference effects when radial velocity (RV) and activity indicator observations are available for a target over multiple observing seasons. We apply our method to two exoplanets discovered by the ${\it K2}$ mission, K2-79b and K2-222b, whose orbital periods overlap with observed activity signals from their host stars. K2-79b was first detected and validated by \citet{crossfield16}, and then re-detected by \citet{Mayo18} and \citet{kruse19}. K2-222b was detected as a planet candidate by \citet{Mayo18} and \citet{petigura18}. It was validated by \citet{Mayo18} and then later re-detected by \citet{Livingston18}. Given their radii of 4.09$\rearth$ and 2.35$\rearth$, respectively, measuring masses for these two planets is particularly interesting, since they lie in the sub-Saturnian desert and near the upper edge of the radius valley where their compositions could be water-rich or water-poor. 


In Section \ref{obs}, we describe the data sets utilized in our analysis, and in section \ref{starchar}, we detail the characterization of the two host stars. In section \ref{act_anal} we describe our investigations of potential stellar activity signals in the light curve (LC) and RV data sets of each target. Section \ref{simult_fit} describes our MCMC fits to determine the exoplanet parameters. We report our results and discuss the scientific implications for K2-79b and K2-222b in Section \ref{discussion}.


\section{Data}\label{obs}

\subsection{K2 Photometry}\label{K2obs}

\begin{figure*}
\subfloat[K2-79]{
\begin{minipage}[c][1\width]{0.45\textwidth}
  \centering
  \includegraphics[width=1\textwidth]{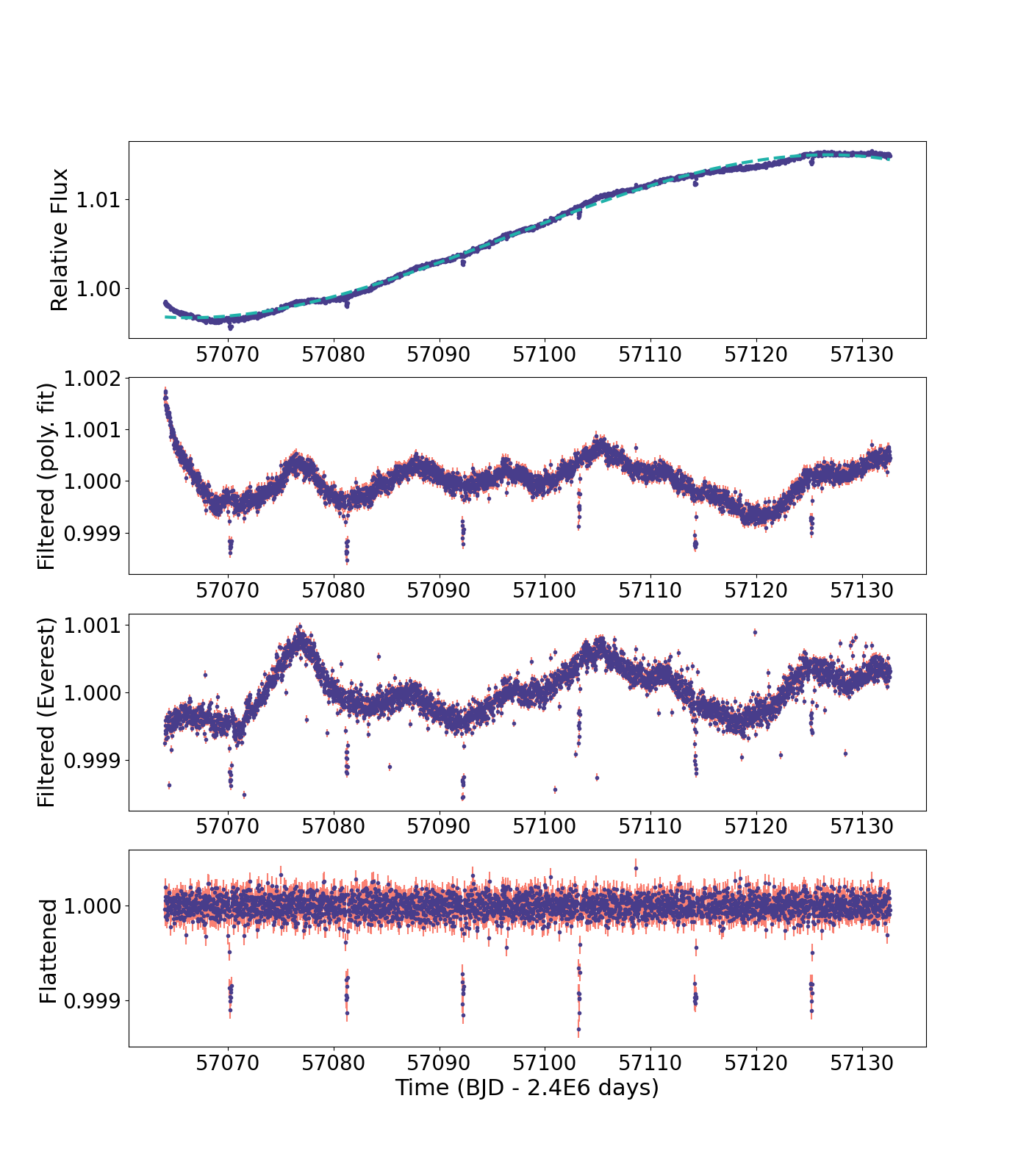}
  \label{2237_K2LC}
\end{minipage}}
\hfill
\subfloat[K2-222]{
\begin{minipage}[c][1\width]{0.45\textwidth}
  \centering
  \includegraphics[width=1\textwidth]{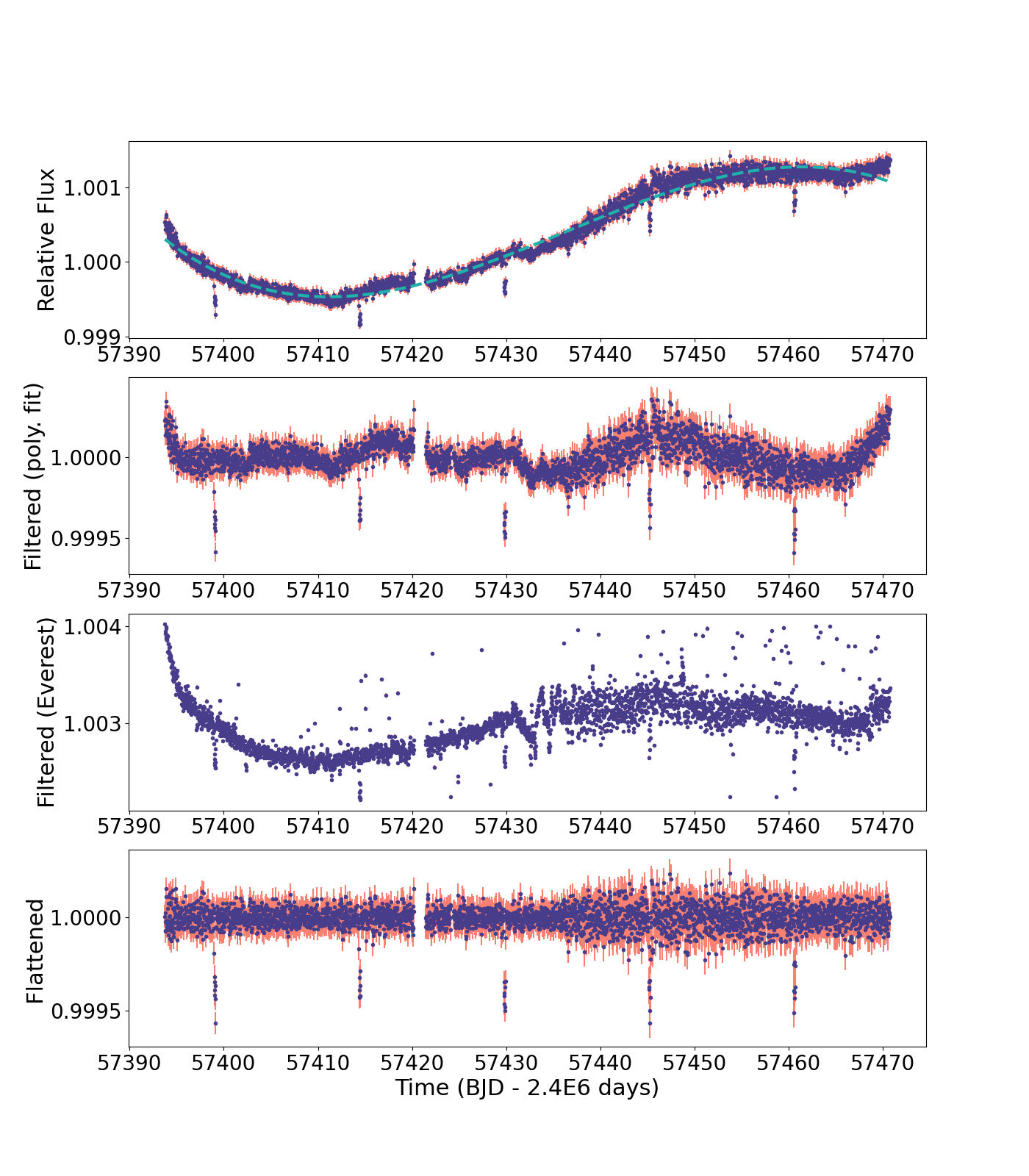}
  \label{9978_K2LC}
\end{minipage}}
\caption{K2 LCs for K2-79 (${\it left}$) and K2-222 (${\it right}$). The top panels show the LCs after removing the thruster firings and other high-frequency systematic signals. The teal dashed lines show the best-fit third-degree polynomials, and the second panels show the LCs after subtracting the respective polynomials (discussed in Section \ref{lc_act_anal}). The third panels show the {\tt Everest}-reduced LCs (also discussed in Section \ref{lc_act_anal}). The bottom panels show the flattened LCs after removing all remaining low-frequency signals. In the case of K2-222, both the scatter of flux values and related errors increase in a later sub-set of the data, beginning at approximately BJD-2400000 = 57438d.}
\label{LCs}
\end{figure*}

We analyzed 3161 and 3424 photometric K2 observations of K2-79 and K2-222, respectively, collected in long cadence (29.4 minutes) mode (Figures \ref{2237_K2LC} and \ref{9978_K2LC}). K2-79 was observed between February 10, 2015 and April 20, 2015, and K2-222 between January 6, 2016 and March 23, 2016. The K2 spacecraft operated with only two of the four original reaction wheels, causing the spacecraft to drift over time, with intermittent thruster firings to keep desired targets in the telescope's field of view. This led to long-term trends in the light curve (LC) of a given star, due to the stars drifting over time across pixels with different quantum efficiencies. We corrected for thruster firings first, followed by systematic instrumental effects and other low frequency signals, according to the technique developed by \citet{vanderburg&johnson2014}. The top panels in Figures \ref{2237_K2LC} and \ref{9978_K2LC} show the LCs of each target after correcting for those effects. After deriving a first-pass systematics corrected light curve, we refined the correction by simultaneously fitting for spacecraft systematics, low-frequency variability, and exoplanet transits following \citet{vanderberg2016}. The bottom panels in the same figures show the resulting flattened LCs with exoplanet transits isolated. We set the error of each point in a given LC data set equal to the standard deviation of the out-of-transit points in the flattened LC. In the case of the brighter star, K2-222, the second half of the observations have increased scatter due to uncorrected K2 systematics. We therefore calculate errors for that LC in two segments separated at BJD - 2400000 = 57438 days, where the start of the increased scatter begins (see Figure \ref{9978_K2LC}, bottom  panel).

\begin{table*}[]
\caption{K2-79 LC observations. \emph{Relative Flux} refers to the LC after removal of K2 thruster firings and other high-frequency systematic signals (Figure \ref{2237_K2LC_pgrams}, top panel). \emph{Flat Relative Flux} refers to the light curve after removal of remaining low-frequency signals (Figure \ref{2237_K2LC_pgrams}, bottom panel). Table \ref{2237LCtab} is published in its entirety in the machine-readable format. A portion is shown here for guidance regarding its form and content.}
\label{2237LCtab}
\centering
\begin{tabular}{lccr}
\hline
Time (BJD - 2.4E6) & Relative flux & Flat Relative Flux & Flux Error \\ 
\hline
57064.0480 & 0.998319 & 0.999963 & 9.550e-05 \\
57064.0680 & 0.998426 & 1.000096 & 9.550e-05 \\
57064.0880 & 0.998189 & 0.999885 & 9.550e-05 \\
57064.1090 & 0.998339 & 1.000060 & 9.550e-05 \\
57064.1290 & 0.998453 & 1.000200 & 9.550e-05 \\
57064.1500 & 0.998322 & 1.000094 & 9.550e-05 \\
57064.1700 & 0.998086 & 0.999882 & 9.550e-05 \\
57064.1910 & 0.998145 & 0.999965 & 9.550e-05 \\
57064.2110 & 0.998011 & 0.999855 & 9.550e-05 \\
57064.2310 & 0.998159 & 1.000027 & 9.550e-05 \\
... & ... & ... & ... \\
\hline
\end{tabular}
\end{table*}

\begin{table*}[]
\caption{K2-222 LC observations. \emph{Relative Flux} refers to the LC after removal of K2 thruster firings and other high-frequency systematic signals (Figure \ref{9978_K2LC_pgrams}, top panel). \emph{Flat Relative Flux} refers to the light curve after removal of remaining low-frequency signals (Figure \ref{9978_K2LC_pgrams}, bottom panel). Table \ref{9978LCtab} is published in its entirety in the machine-readable format. A portion is shown here for guidance regarding its form and content.}
\label{9978LCtab}
\centering
\begin{tabular}{lccr}
\hline
Time (BJD - 2.4E6) & Relative Flux & Flat Relative Flux & Relative Flux Error \\
\hline
57393.7440 & 1.000534 & 1.000005 & 6.061e-05 \\
57393.7650 & 1.000501 & 0.999979 & 6.061e-05 \\
57393.7850 & 1.000541 & 1.000025 & 6.061e-05 \\
57393.8050 & 1.000435 & 0.999925 & 6.061e-05 \\
57393.8260 & 1.000434 & 0.999931 & 6.061e-05 \\
57393.8460 & 1.000608 & 1.000112 & 6.061e-05 \\
57393.8870 & 1.000637 & 1.000153 & 6.061e-05 \\
57393.9080 & 1.000446 & 0.999967 & 6.061e-05 \\
57393.9280 & 1.000455 & 0.999983 & 6.061e-05 \\
57393.9480 & 1.000417 & 0.999951 & 6.061e-0 \\
... & ... & ... & ... \\
\hline
\end{tabular}
\end{table*}


\subsection{HARPS-N Spectroscopy}

\begin{figure*}[]
\epsscale{1.05} 
\plotone{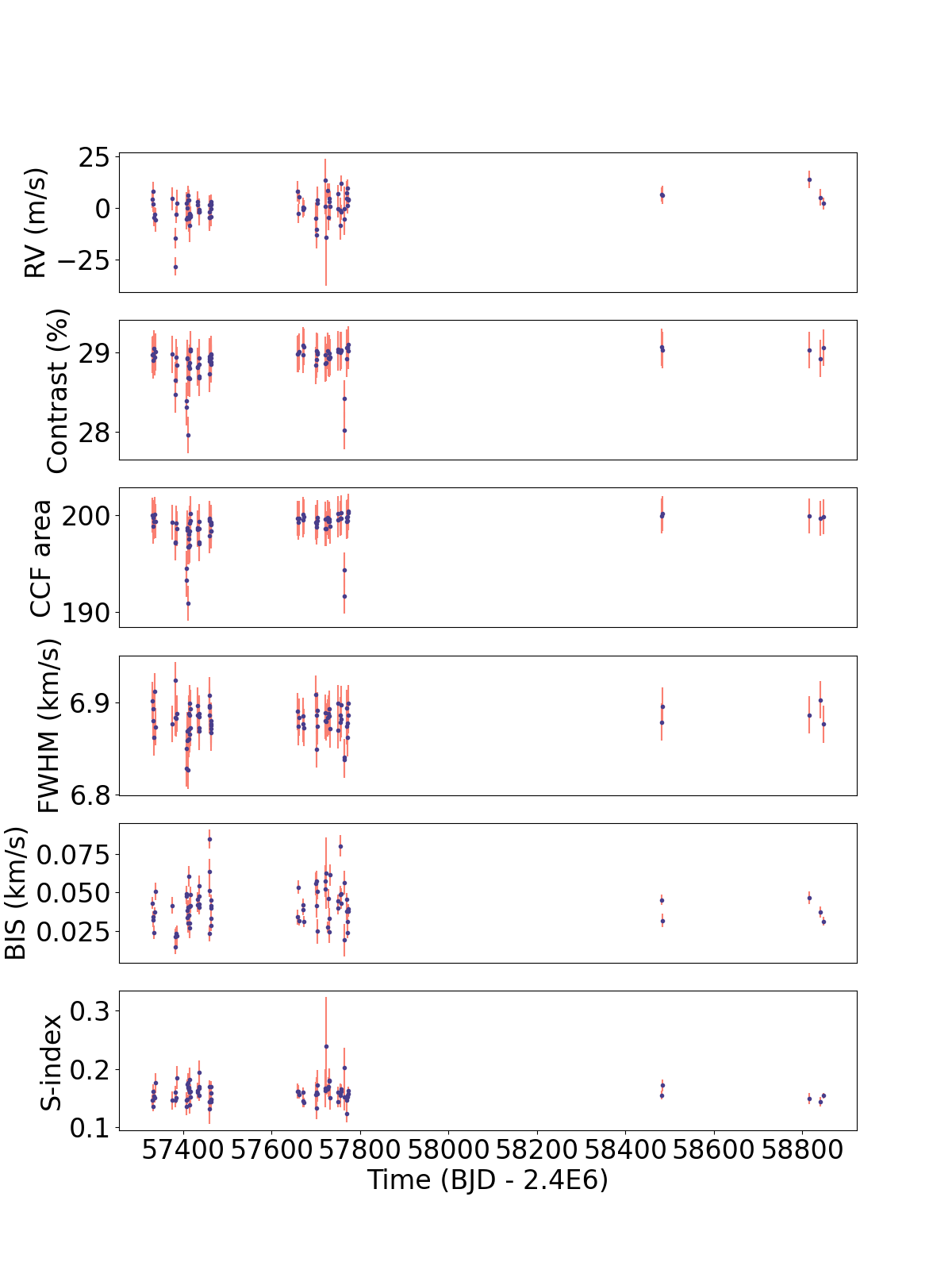}
\caption{K2-79 HARPS-N RV and stellar activity indicator (contrast, cross-correlation function area, full-width at half maximum, bisector inverse span, and S-index) observations.}\label{2237_RVscat}
\end{figure*}

\begin{figure*}[]
\epsscale{1.05} 
\plotone{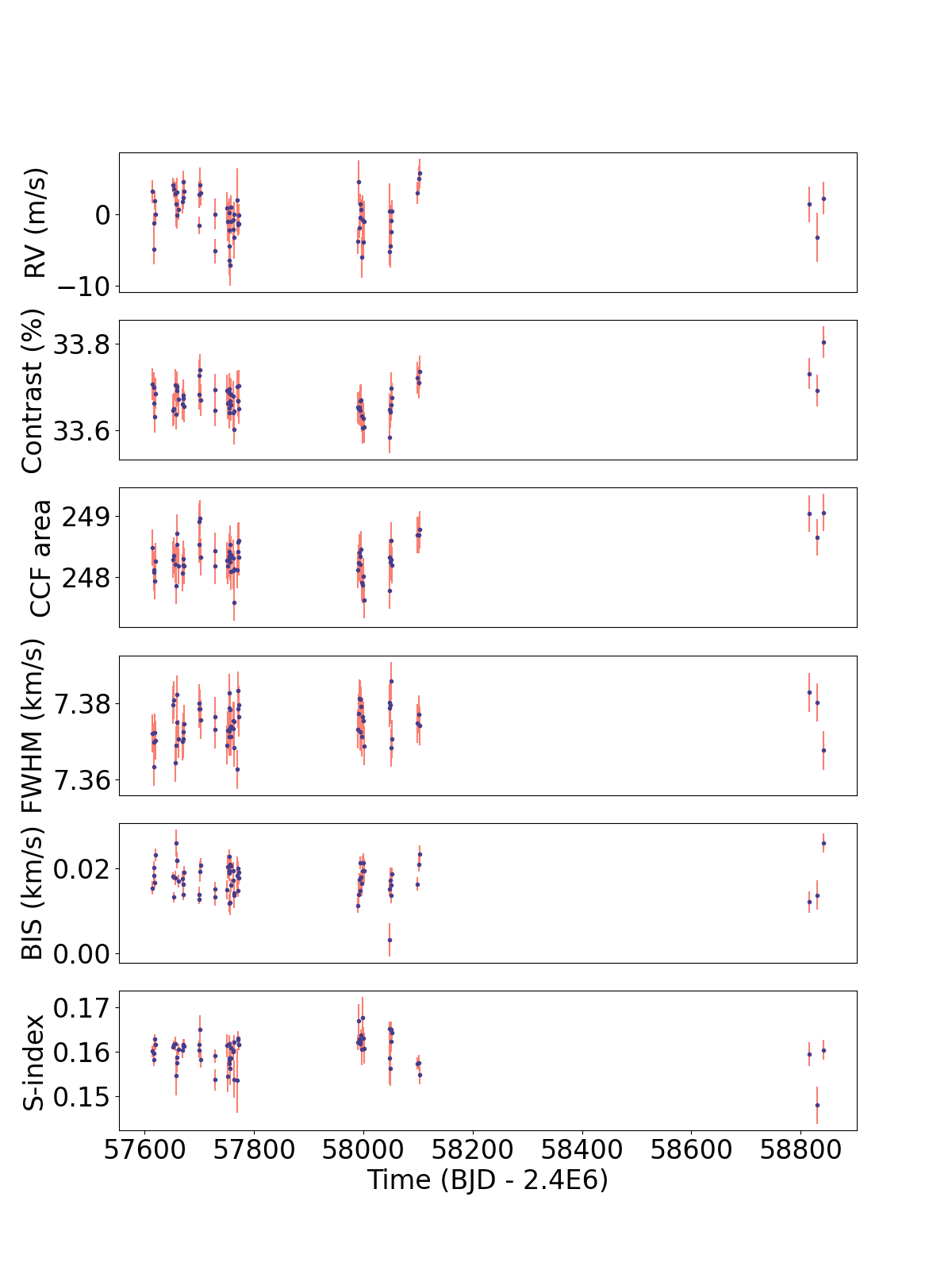}
\caption{K2-222 HARPS-N RV and stellar activity indicator (contrast, cross-correlation function area, full-width at half maximum, bisector inverse span, and S-index) observations.}\label{9978_RVscat}
\end{figure*}

We collected 79 spectra of K2-79 over four seasons between November 4, 2015 and December 29, 2019 (Figure \ref{2237_RVscat}), and 63 spectra of K2-222 over three seasons between August 14, 2016 and December 23, 2019 (Figure \ref{9978_RVscat}). All spectra were collected with HARPS-N, the high precision spectrograph mounted on the Telescopio Nationale de Galileo at the Observatorio del Roque de los Muchachos in La Palma, Spain \citep{cosentino12}. The spectrograph covers wavelengths in the range 383 - 690 nm, with average resolution $R = $ 115,000. Observations for K2-79 were taken with 30 minute exposure times, yielding a mean signal-to-noise ratio (SNR) of 32.67 in order 60 (627.077 - 634.166 nm) of the spectrum. For K2-222, observation exposure times ranged between 15 - 30 minutes, yielding a mean SNR of 74.11 in order 60. The observations were collected as part of the HARPS-N collaboration's Guaranteed Time Observations (GTO) program.

We reduced the spectra and calculated associated RVs and activity indicators using the HARPS-N data reduction software \citep[DRS;][]{lovis&pepe2007}. Activity indicators calculated include the contrast, FWHM, cross-correlation function area (CCF area = contrast * FWHM), bisector inverse span (BIS), and S-index (Figures \ref{2237_RVscat} and \ref{9978_RVscat}). 

\begin{table}[]
    \centering
    \caption{Number and time range of observed HARPS-N RV and activity indicator data sets, as well as full seasonal sub-sets.} \label{Nobs}
    \begin{tabular}{lccr}
        \hline
         & Full & S1 & S2 \\
        \hline
         K2-79 & & & \\
         N$_{obs}$ & 78 & 40 & 33 \\
         Obs. Baseline (days) & 1517 & 133 & 114 \\
         
         K2-222 & & & \\
         N$_{obs}$ & 63 & 41 & 19 \\
         Obs. Baseline (days) & 1227 & 158 & 114 \\
        \hline
    \end{tabular}
\end{table}

\begin{table*}[]
\caption{K2-79 HARPS-N RV and stellar activity indicator (contrast, cross-correlation function (CCF) area, full-width at half maximum (FWHM), bisector inverse span (BIS), and S-index) observations (Figure \ref{2237_RVscat}). Table \ref{2237RVtab} is published in its entirety in the machine-readable format. A portion is shown here for guidance regarding its form and content.}
\label{2237RVtab}
\hskip-1.5cm
\begin{tabular}{lcccccccr}
\hline
Time (BJD - 2.4e6) & RV (km/s) & RV error (km/s) & FWHM (km/s) & Contrast (\%) & BIS (km/s) & BIS error (km/s) & S-index & S-index error \\ 
\hline
57330.533 & 4.82 & 4.09 & 6.902 & 28.972 & 0.04314 & 0.00406 & 0.14638 & 0.00983 \\
57331.519 & 2.42 & 3.65 & 6.893 & 28.977 & 0.03406 & 0.00361 & 0.13526 & 0.00829 \\
57332.517 & 8.54 & 4.72 & 6.880 & 28.903 & 0.03194 & 0.00470 & 0.16132 & 0.01230 \\
57333.561 & -3.81 & 4.40 & 6.862 & 29.048 & 0.02392 & 0.00437 & 0.15386 & 0.01165 \\
57334.686 & -2.46 & 3.47 & 6.911 & 28.942 & 0.03705 & 0.00343 & 0.15081 & 0.00778 \\
57336.468 & -5.19 & 5.65 & 6.873 & 29.005 & 0.05062 & 0.00563 & 0.17573 & 0.01690 \\
57374.538 & 5.06 & 5.59 & 6.877 & 28.976 & 0.04152 & 0.00557 & 0.14590 & 0.01614 \\
57380.443 & -41.96 & 7.55 & 6.971 & 28.108 & -0.03486 & 0.00754 & 0.16431 & 0.02248 \\ 
57381.538 & -27.68 & 4.50 & 6.924 & 28.474 & 0.01424 & 0.00448 & 0.15965 & 0.01113 \\
57382.471 & -14.06 & 4.98 & 6.884 & 28.647 & 0.02122 & 0.00496 & 0.14700 & 0.01299 \\
 ... & ... & ... & ... & ... & ... & ... & ... & ... \\
\hline
\end{tabular}
\end{table*}


\begin{table*}[]
\caption{K2-222 HARPS-N RV and stellar activity indicator (contrast, cross-correlation function (CCF) area, full-width at half maximum (FWHM), bisector inverse span (BIS), and S-index) observations (Figure \ref{9978_RVscat}). Table \ref{9978RVtab} is published in its entirety in the machine-readable format. A portion is shown here for guidance regarding its form and content.}
\label{9978RVtab}
\hskip-1.5cm
\begin{tabular}{lcccccccr}
\hline
Time (BJD - 2.4e6) & RV (km/s) & RV error (km/s) & FWHM (km/s) & Contrast (\%) & BIS (km/s) & BIS error (km/s) & S-index & S-index error \\
\hline
57614.698 & 3.43 & 1.58 & 7.372 & 33.706 & 0.01540 & 0.00154 & 0.16018 & 0.00137 \\
57616.700 & -4.64 & 2.17 & 7.370 & 33.662 & 0.01839 & 0.00214 & 0.15968 & 0.00239 \\
57617.707 & -1.01 & 1.53 & 7.363 & 33.698 & 0.02025 & 0.00149 & 0.15820 & 0.00132 \\
57618.707 & 2.19 & 1.32 & 7.372 & 33.631 & 0.01670 & 0.00127 & 0.16294 & 0.00102 \\
57619.740 & 0.24 & 1.49 & 7.370 & 33.685 & 0.02321 & 0.00145 & 0.16168 & 0.00125 \\
57652.507 & 4.34 & 1.00 & 7.380 & 33.645 & 0.01821 & 0.00093 & 0.16119 & 0.00065 \\
57653.511 & 3.86 & 1.24 & 7.381 & 33.649 & 0.01319 & 0.00119 & 0.16163 & 0.00097 \\
57655.675 & 3.15 & 1.71 & 7.364 & 33.705 & 0.01768 & 0.00168 & 0.16181 & 0.00164 \\
57657.610 & 1.65 & 3.21 & 7.369 & 33.637 & 0.02601 & 0.00319 & 0.15475 & 0.00442 \\
57658.675 & 3.32 & 2.05 & 7.375 & 33.700 & 0.02196 & 0.00202 & 0.15761 & 0.00213 \\
 ... & ... & ... & ... & ... & ... & ... & ... & ... \\
\hline
\end{tabular}
\end{table*}


\section{Stellar Characterization}\label{starchar}

We derived stellar parameters for each target using the Stellar Parameter Classification Tool \citep[SPC;][]{Buchhave12, Buchhave14} to analyze the K2-79 and K2-222 HARPS-N spectra and estimate stellar effective temperature (T$_{\rm eff}$), surface gravity ($\logg$), metallicity ($\rm [Fe/H]$), and projected rotational velocity ($\vsini$) for each target. Spectra with normalized peak heights of less than 0.9 in the cross-correlation function were discarded due to low signal-to-noise ratios. We aligned all individual spectra to a common RV frame and calculated the maximum stellar rotation period associated with the lower limit of $\vsini$ using the equation $P_{\rm{rot, max}} = 2\pi R_{*} / \vsini$. Final values from SPC are summarized in Table \ref{starpars}. 

We also applied the ARES+MOOG equivalent width method to the K2-79 and K2-222 HARPS-N spectra to calculate T$_{\rm eff}$, $\logg$, $\rm [Fe/H]$, and microturbulence ($\chi_{\rm turb}$) \citep{Sousa14} for each target. Spectra associated with 3-sigma RV outliers were discarded. As with SPC, all used spectra were shifted to a common RV frame before combining them for analysis. For accuracy, the value of $\logg$ was corrected following Equation 3 in \citet{Mortier14}, and its error bars were increased according to \citet{Sousa11}. Final values measured with ARES+MOOG and the combined results with SPC are summarized in Table \ref{starpars}.  

Finally, with the parallax for each star from the Gaia Early Data Release 3 \citep{DR3} and stellar magnitudes in various filters (B,V,J,H,K,W1,W2,W3, details in Table \ref{starpars}), we used the {\tt Isochrones} \citep{isochrones} software to run four isochrone analyses for each target: Dartmouth models \citep{Dartmouth} and MIST models \citep{MIST} each paired with both the SPC and ARES + MOOG obtained T$_{\rm eff}$ and metallicity values. The Gaia parallax and magnitudes in various filter are listed in Table \ref{starpars}. We used 400 live points for the \textit{Multinest} algorithm and combined posteriors from all four isochrone runs, as first presented in \citet{Luca2018}, to obtain final values for T$_{\rm eff}$, $\logg$, $\rm [Fe/H]$, extinction (A(V)), stellar luminosity, radius, mass, density, distance, and age. For both targets, parameter values between our four isochrone runs were in agreement with each other within 1.5-sigma. The final isochrone values listed in Table \ref{starpars} were obtained by combining results from the four isochrone fits. When applicable, we inflate errors on values obtained with any method to 2\% in temperature and radius, 4\% in luminosity, 5\% in mass, and 20\% in age, according to \citet{Tayar2020}'s approximations for the systematic uncertainty floor associated with current measured parallaxes and bolometric fluxes. The final derived isochrone results had 1-sigma agreement with relevant parameters measured in our SPC and ARES+MOOG analyses, as well as those previously measured by \citet{huber2016}.

\begin{table*}[]
  \caption{Stellar parameters for K2-79 and K2-222. For values with multiple estimates, we indicate the adopted value with asterisks. Abbreviations for analyses applied to the HARPS-N spectra are: A+M = ARES + MOOG \citep{Sousa14}, SPC = Stellar Parameter Classification tool \citep{Buchhave12, Buchhave14}. The adopted isochrone values were obtained by combining results from four different isochrone fits detailed in Section \ref{starchar}. Tables \ref{exofit_2237_tab} and \ref{exofit_9978_tab} list stellar parameters for K2-79 and K2-222 obtained with fits using the {\tt EXOFAST v2} software \citep{eastman19}.} \label{starpars}
         \small
         \centering
   \begin{tabular}{lccr}
            \hline
            \noalign{\smallskip}
            Parameter  & K2-79 & K2-222 & Source \\
            \noalign{\smallskip}
            \hline
            \noalign{\smallskip}
            EPIC ID & 210402237 & 220709978 &  \\
            RA [deg] & 55.255896 & 16.462277 & \citet{DR3} \\
            DEC [deg] & 13.519365 & 11.753428 & '' \\
            Parallax [mas] & 3.841 $\pm$ 0.015 & 10.00896 $\pm$ 0.06080 & '' \\
            G & 11.799 $\pm$ 0.0003 & 9.3659 $\pm$ 0.0003 & '' \\
            B & 12.89 $\pm$ 0.02 & 10.04 $\pm$ 0.03 & \citet{zacharias2012} \\
            V & 12.07 $\pm$ 0.06 & 9.54 $\pm$ 0.03 & '' \\
            J & 10.36 $\pm$ 0.02 & 8.42 $\pm$ 0.02 & \citet{cutri2003} \\
            H & 10.00 $\pm$ 0.02 & 8.17 $\pm$ 0.03 & '' \\
            K & 9.91 $\pm$ 0.02 & 8.11 $\pm$ 0.02 & '' \\
            W1mag & 9.83 $\pm$ 0.02 & 8.05 $\pm$ 0.02 & \citet{cutri2013} \\
            W2mag & 9.86 $\pm$ 0.02 & 8.10 $\pm$ 0.02 & '' \\
            W3mag & 9.80 $\pm$ 0.06 & 8.07 $\pm$ 0.02 & '' \\
            
            Reddening (G-R) & 0.23$^{+0.03}_{-0.02}$ & 0.0 $^{+0.01}_{-0.0}$ & \citet{green19} \\
            Reddening (B-V) & 0.20$^{+0.03}_{-0.02}$ & 0.0 $^{+0.01}_{-0.0}$ & \citet{green19} \\
            Extinction, A(V) & 0.81$^{+0.06}_{-0.09}$ & 0.03$^{+0.05}_{-0.02}$ & isochrones \\

            Distance [pc] & 260 $\pm$ 1 & 99.6 $\pm$ 0.2 & isochrones \\
            Mass [$\msun$] & 1.06 $\pm$ 0.05 & 0.94$\pm$ 0.05 & isochrones* \\
            &0.992 $\pm$ 0.103 & 1.054 $\pm$ 0.104 & \citet{huber2016} \\
            Radius [$\rsun$] & 1.269 $\pm$ 0.051  & 1.072 $\pm$ 0.043 & isochrones* \\
            & 1.1 $\pm$ 0.4 & 1.1 $\pm$ 0.5 & \citet{huber2016}\\
            Density [$\rhosun$] & 0.52 $\pm$ 0.02 & 0.76$^{+0.05}_{-0.04}$ & isochrones* \\
            & 0.75$\pm$0.30 & 0.79 $\pm$ 0.30 & \citet{huber2016} \\
            Luminosity [$\lsun$] & 1.76$^{+0.06}_{-0.07}$ & 1.34 $\pm$ 0.04 & isochrones \\
            Age [Gyr] & 6.5 $\pm$ 1.3 & 7.1$^{+1.5}_{-1.7}$ & isochrones \\
            Eff. temp., T$_{\rm eff}$ [K] & 5897 $\pm$ 118 & 5942 $\pm$ 119 & Combined - A+M \& SPC* \\
            & 5901 $\pm$ 118 & 6002 $\pm$ 120 & isochrones \\
            & 5943 $\pm$ 119  & 6000 $\pm$ 120 & A+M \\
            & 5851 $\pm$ 117 & 5883 $\pm$ 118 & SPC \\
            & 5926 $\pm$ 86 & 5818 $\pm$ 136  & \citet{huber2016} \\
            Metallicity, $\rm [Fe/H]$ & 0.035 $\pm$ 0.06 & -0.315 $\pm$ 0.06 & Combined - A+M \& SPC* \\
            & 0.05$^{+0.05}_{-0.04}$ & -0.25$^{+0.04}_{-0.11}$ & isochrones \\
            & 0.04 $\pm$ 0.04 & -0.26 $\pm$ 0.04 & A+M \\
            & 0.03 $\pm$ 0.08 & -0.37 $\pm$ 0.08 & SPC \\
            Surface grav, $\logg$ [cgs] & 4.26 $\pm$ 0.01 & 4.35 $\pm$ 0.02 & isochrones* \\
            & 4.21 $\pm$ 0.11 & 4.24 $\pm$ 0.12 & A+M \\
            & 4.29 $\pm$ 0.1 & 4.19 $\pm$ 0.1 & SPC \\
            Microturb., $\xi_{\rm turb}$ [\kms] & 1.09$\pm$0.04 & 1.14 $\pm$ 0.04 & A+M \\
            v$\sin$i [\kms] & 2.7 $\pm$ 0.5 & 3.1 $\pm$ 0.5 & SPC \\
            P$_{\rm rot, max}$ [days] & 23.8 $\pm$ 4.4 & 17.5 $\pm$ 2.8 & SPC (calculated from v$\sin$i) \\
            
            \noalign{\smallskip}
            \hline
     \end{tabular}  
\end{table*}


\section{Stellar Activity Analysis}\label{act_anal}

We performed analyses of the K2-79 and K2-222 LCs and the HARPS-N RVs and activity indicators to search for prominent stellar activity signals. The next two subsections detail these analyses.

\subsection{Light Curve Stellar Activity Analysis}\label{lc_act_anal}

\begin{figure*}
\caption{K2 LCs for K2-79 (${\it left}$) and K2-222 (${\it right}$). The top panels show the LCs after removing the thruster firings and other high-frequency systematic signals. The teal dashed lines show the best-fit third-degree polynomials, and the second panels show the LCs after subtracting the respective polynomials (discussed in Section \ref{lc_act_anal}). The third panels show the {\tt Everest}-reduced LCs (also discussed in Section \ref{lc_act_anal}). The bottom panels show the flattened LCs after removing all remaining low-frequency signals. In the case of K2-222, both the scatter of flux values and related errors increase in a later sub-set of the data, beginning at approximately BJD-2400000 = 57438d.}
\subfloat[K2-79: After removing the best-fit third-order polynomial (second panel), additional statistically significant peaks occur at 22.9d (f = 0.0437$d^{-1}$), 16.4d (f = 0.0610$d^{-1}$), 7.43d (f = 0.1345$d^{-1}$), 6.667d (f = 0.1500$d^{-1}$), and 5.999d (f = 0.1667$d^{-1}$). After applying the {\tt Everest} reduction (third panel), additional statistically significant peaks occur at 25.9d (f = 0.0386$d^{-1}$), 16.7d (f = 0.0599$d^{-1}$), and 7.0d (f = 0.1427$d^{-1}$). The strongest peak in each is consistent with the estimated $P_{\rm rot,max}$ (23.8d $\pm$ 4.4d) within 0.5$\sigma$, and may be related to stellar rotation.]{
\begin{minipage}[t][1\width]{0.45\textwidth}
  \centering
  \includegraphics[width=1\textwidth]{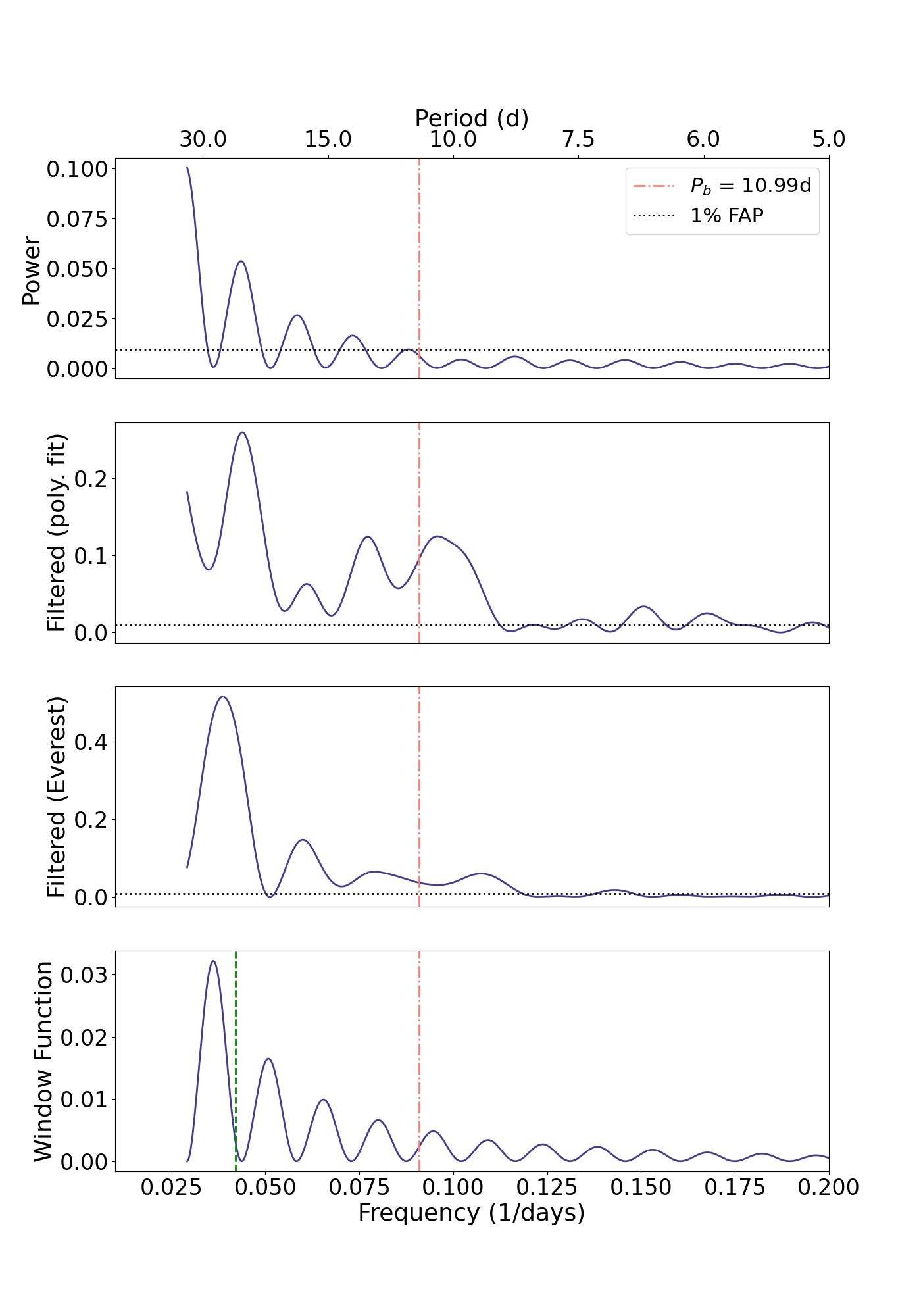}
  \label{2237_K2LC_pgrams}
\end{minipage}}
\hfill
\subfloat[K2-222: After removing the best-fit third-order polynomial (second panel), additional statistically significant peaks occur at 28.2d (f = 0.0355$d^{-1}$) and 7.6d (f = 0.1316$d^{-1}$). After applying the {\tt Everest} reduction (third panel), additional statistically significant peaks occur at 25.7d (f = 0.0389$d^{-1}$), 18.8d (f = 0.0532$d^{-1}$), 15.2d (f = 0.0658$d^{-1}$), 8.4d (f = .1190$d^{-1}$), and 6.8d (f = 0.1471$d^{-1}$). The strongest peaks at 28.2d and 25.7d in the polynomial- and {\tt Everest}-reduced LCs, respectively, are barely within 4-sigma and 3-sigma of the estimated $P_{\rm rot,max}$ (17.5d $\pm$ 4.4d). Other pronounced peaks unrelated to stellar rotation estimates could still be related to stellar activity and are investigated further in RVs.]{
\begin{minipage}[t][1\width]{0.45\textwidth}
  \centering
  \includegraphics[width=1\textwidth]{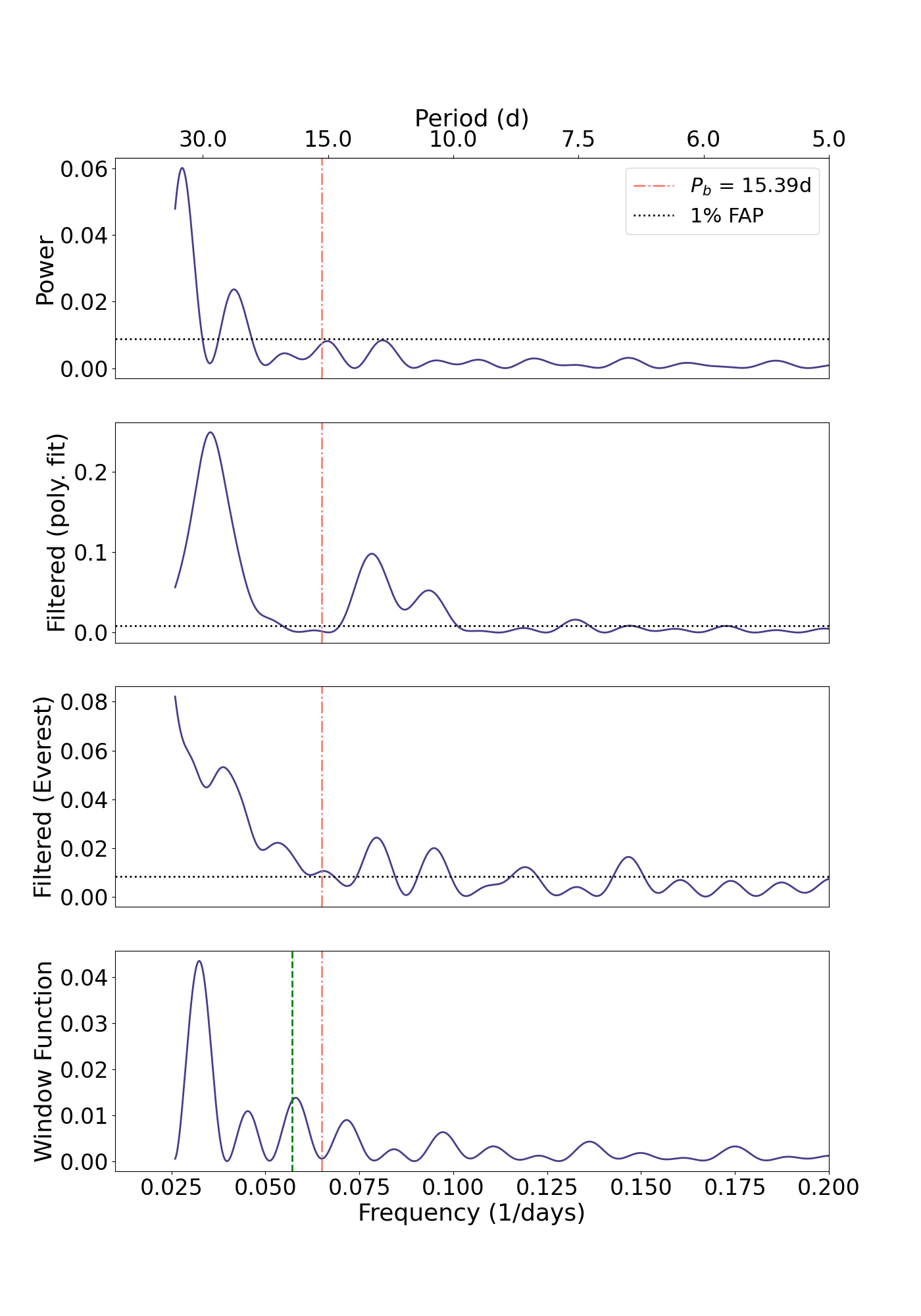}
  \label{9978_K2LC_pgrams}
\end{minipage}}
\label{LC_pgrams}
\end{figure*}

To probe prominent stellar activity signals, we calculated the Generalized Lomb-Scargle (GLS) periodograms \citep{Zech2009} on the LC of each target with only K2 thruster firings removed (top panels in Figure \ref{LCs}). We also performed auto-correlation function (ACF) analysis on each LC. We utilized the {\tt Astropy.timeseries Lomb-Scargle} package to calculate all GLS periodograms using the default {\it auto} method which selects the best periodogram algorithm based on the input data. We calculate all periodograms with a sampling of N = 50,000 for periods between 0.1 days and one-half the baseline time span of the LC, or corresponding frequencies \citep{Astropy_a, Astropy_b}. We calculate a 1$\%$ false alarm probability (FAP) level using the {\tt Astropy.timeseries Lomb-Scargle} package's default {\it false\_alarm\_level} tool, which calculates the upper-limit to the alias-free probability using the approach outlined in \citet{baluev2008}. The resulting periodograms of the transit-\emph{masked} K2-79 and K2-222 LCs are shown in the top panels of Figure \ref{LC_pgrams}.

To calculate each \emph{window function}, we produce a signal with the same time stamps as the observations, replacing the LC fluxes with ones. We then calculate a GLS periodogram on the signal with the same period and frequency limits used to calculate the LC periodogram, this time applying the {\it scipy} periodogram algorithm which treats individual points without error. The resulting window functions of the K2-79 and K2-222 LCs are shown in the bottom panels of Figure \ref{LC_pgrams}. 

In the cases of both targets, the top panel periodograms are dominated by long-term instrumental systematics that can be seen by eye in the LCs in the top panels of Figure \ref{LCs}. Therefore, for each target, we perform periodogram analyses on two more LCs: one with the best-fit third-degree polynomial subtracted (second panels in Figures \ref{LCs} and \ref{LC_pgrams}) and the other reduced with {\tt Everest} \citep{Luger16}, an open-source pipeline for removing instrumental noise signals from K2 LCs (third panels in Figures \ref{LCs} and \ref{LC_pgrams}). We truncated the Everest LCs to match the baseline of the observations used in our analysis and removed extreme outliers. After these cuts, the {\tt Everest}-reduced LCs still include some points excluded in our reduction, but the window functions for the two sets are identical.

Two statistically significant (FAP $<$ 1$\%$) signals found in both targets' periodograms are likely due to remaining high-frequency instrumental systematics in the LCs. In the K2-79 and K2-222 polynomial- and {\tt Everest}-reduced LCs, respectively, the first signal occurs at 12.9d, 12.7d, 12.9d, and 12.6d, and the second at 10.3d, 9.33d, 10.7d and 10.6d. We attribute the scatter in values to the different reduction methods being applied, but the set of peaks can be spotted by-eye in both targets' periodograms.

In the case of K2-79, after removing the best-fit third-order polynomial, additional statistically significant peaks occur in the LC periodogram at 22.9d (f = 0.0437$d^{-1}$), 16.4d (f = 0.0610$d^{-1}$), 6.7d (f = 0.1500$d^{-1}$), and 6.0d (f = 0.1667$d^{-1}$) (Figure \ref{2237_K2LC_pgrams}, second panel). After applying the {\tt Everest} reduction, additional statistically significant peaks occur at 25.9d (f = 0.0386$d^{-1}$), 16.7d (f = 0.0599$d^{-1}$), and 7.0d (f = 0.1427$d^{-1}$) (Figure \ref{2237_K2LC_pgrams}, third panel). The strongest peaks at 22.9d and 25.9d in the polynomial- and {\tt Everest}-reduced LCs, respectively, are consistent with the estimated $P_{\rm rot,max}$ (23.8d $\pm$ 4.4d) within 0.5$\sigma$, and may be related to stellar rotation. In the case of K2-222, after removing the best-fit third-order polynomial, additional statistically significant peaks occur at 28.2d (f = 0.0355$d^{-1}$) and 7.6d (f = 0.1316$d^{-1}$) (Figure \ref{9978_K2LC_pgrams}, second panel). After applying the {\tt Everest} reduction, additional statistically significant peaks occur at 25.7d (f = 0.0389$d^{-1}$), 18.8d (f = 0.0532$d^{-1}$), 15.2d (f = 0.0658$d^{-1}$), 8.4d (f = .1190$d^{-1}$), and 6.8d (f = 0.1471$d^{-1}$) (Figure \ref{9978_K2LC_pgrams}, third panel). The strongest peaks at 28.2d and 25.7d in the polynomial- and {\tt Everest}-reduced LCs, respectively, are barely within 4-sigma and 3-sigma of the estimated $P_{\rm rot,max}$ (17.5d $\pm$ 4.4d).

We calculated the ACF by applying discrete shifts to the polynomial-reduced and {\tt Everest}-reduced LCs and cross-correlating the shifted LCs with the original. The ACF shows repeating peaks separated by \emph{$P_{\rm quasi}$}, a timescale often related to \emph{$P_{\rm rot}$}, with the correlation power of each peak dropping off at a rate related to the evolution timescale of magnetic active regions, \emph{$\tau$}. We utilized the {\tt scipy} non-linear least squares function to fit an under-damped simple harmonic oscillator curve to each ACF, described by:
\begin{equation}
    \label{dsho}
    y = e^{-t/\tau} \left[ A\cos{ \left( \frac{2\pi t}{P_{\rm quasi}} \right)} + y_{0} \right],
\end{equation}
where \emph{A} is the ACF amplitude, and y$_{0}$ is the oscillator offset from zero. The fits provide estimates of \emph{$P_{\rm quasi}$} and \emph{$\tau$} for each star \citep{mcquillan14, giles17}. For each fit we also tried a model accounting for the common scenario of two magnetic active regions evenly spaces around the star. We did this be including a second cosine term, $B\cos{ \left(\frac{2\pi t}{P_{\rm quasi}}\right)}$, inside the brackets of Equation \ref{dsho}. However, all of the best fits returned and amplitude consistent with zero for B. We also place an upper bound of 100 days on \emph{$\tau$}, as it will suffice to identify a stable ($\tau >> P_{\rm quasi}$) signal with only 35-40 day ACF baselines.

\begin{figure*}
\caption{Auto-correlation function (ACF) curves for the polynomial-reduced (${\it top}$) and {\tt Everest}-reduced (${\it bottom}$) K2 LCs of K2-79 (${\it left}$) and K2-222 (${\it right}$). The dark purple points show the ACF. The solid orange line shows the result of the under damped simple harmonic oscillator function fit to the ACF curves, as described in Section \ref{lc_act_anal}. The ACF shows repeating peaks separated by \emph{$P_{\rm quasi}$}, a timescale often related to \emph{$P_{\rm{rot}}$}, with the correlation power of each peak dropping off at a rate related to the evolution timescale of magnetic active regions, \emph{$\tau$}. The \emph{P$_{\rm quasi}$} values returned by successful fits for both targets are similar to each other and the strongest signals in all LC periodograms (Figure \ref{LC_pgrams}). The \emph{$\tau$} values are also all quite long relative to \emph{P$_{\rm quasi}$}. We therefore cannot rule out, for either target, the potential that the ACF signal is the result of instrumental systematics.}

\subfloat[K2-79: The fit to the ACF of the polynomial-reduced LC reveals the strongest periodicity at P$_{\rm quasi}$ = 26.6 $\pm$ 0.4 days, with an evolution timescale of \emph{$\tau$} = 72 $\pm$ 6 days. The ACF of the {\tt Everest}-reduced LC reveals the strongest periodicity at P$_{\rm quasi}$ = 26.1 $\pm$ 0.2 days, with an evolution timescale of \emph{$\tau$} = 100 $\pm$ 10 days.]{
\begin{minipage}[t][1\width]{0.45\textwidth}
  \centering
  \includegraphics[width=1\textwidth]{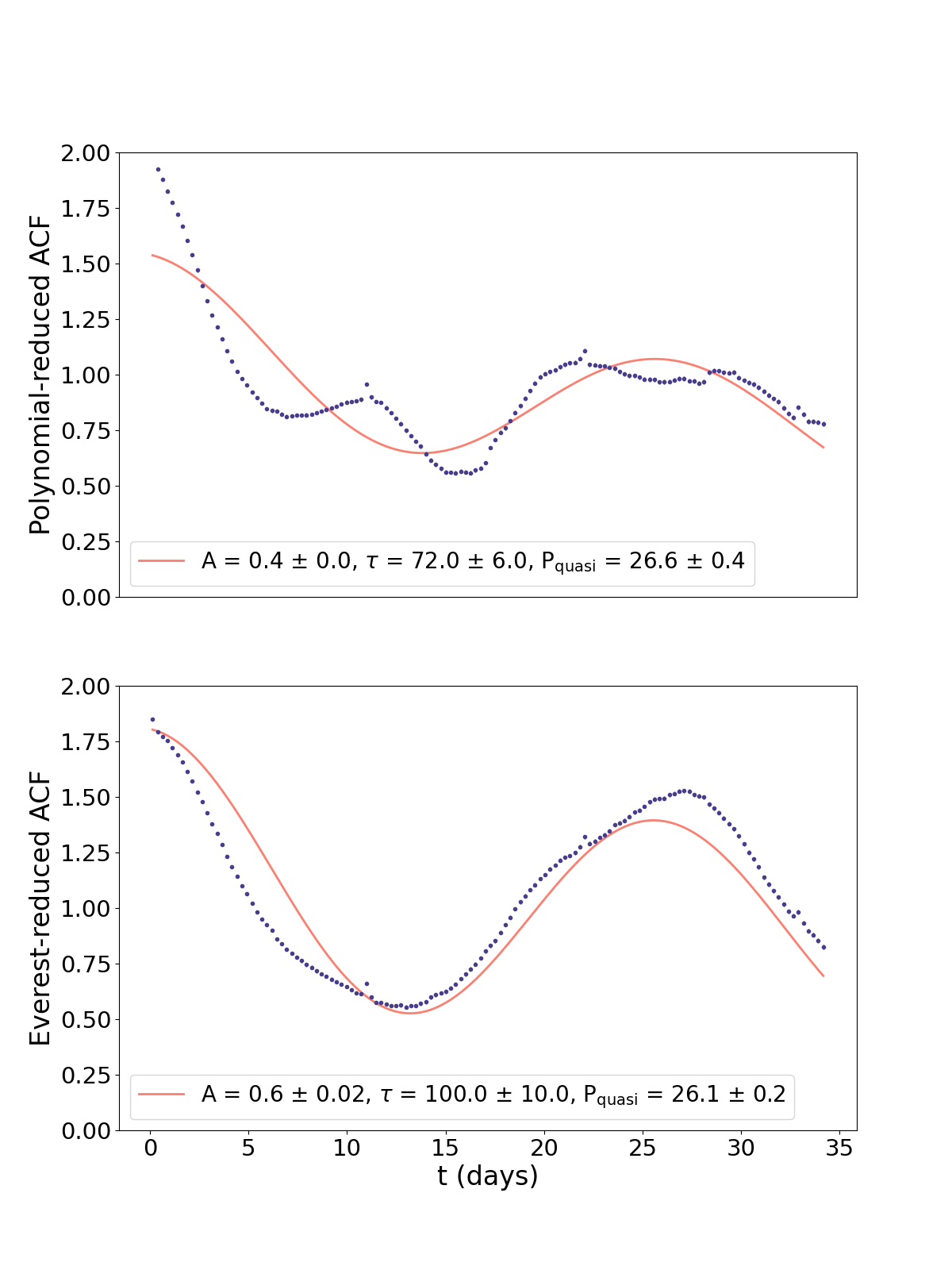}
  \label{2237_ACF}
\end{minipage}}
\hfill
\subfloat[K2-222: The fit to the ACF of the polynomial-reduced LC reveals the strongest periodicity at P$_{\rm quasi}$ = 28.2 $\pm$ 0.2 days, with an evolution timescale of $\tau$ = 100 $\pm$ 10 days. The ACF of the {\tt Everest}-reduced LC returns an amplitude (A) consistent with zero, dominated by the leftover long-period instrumental trend that can be seen by-eye in the LC (Figure \ref{9978_K2LC}, third panel).]{
\begin{minipage}[t][1\width]{0.45\textwidth}
  \centering
  \includegraphics[width=1\textwidth]{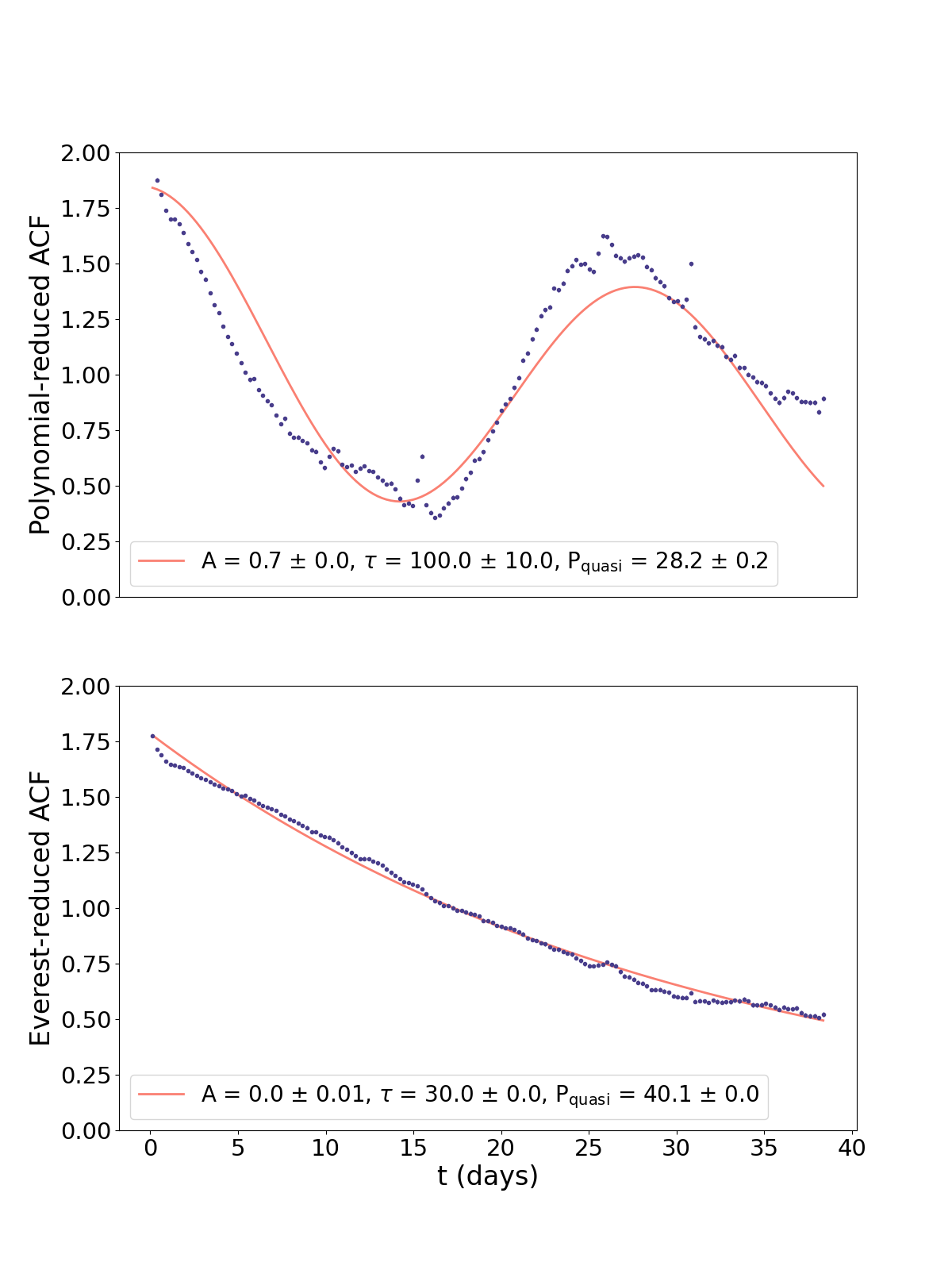}
  \label{9978_ACF}
\end{minipage}}
\label{ACFs}
\end{figure*}

The results of the ACF analysis are shown in Figure \ref{2237_ACF} for the K2-79 LC and in Figure \ref{9978_ACF} for K2-222. In the case of K2-79, the fit to the ACF of the polynomial-reduced LC reveals the strongest periodicity at P$_{\rm quasi}$ = 26.6 $\pm$ 0.4 days, with an evolution timescale of \emph{$\tau$} = 72 $\pm$ 6 days. The ACF of the {\tt Everest}-reduced LC reveals the strongest periodicity at P$_{\rm quasi}$ = 26.1 $\pm$ 0.2 days, with an evolution timescale of \emph{$\tau$} = 100 $\pm$ 10 days. In the case of K2-222, the fit to the ACF of the polynomial-reduced LC reveals the strongest periodicity at P$_{\rm quasi}$ = 28.2 $\pm$ 0.2 days, with an evolution timescale of $\tau$ = 100 $\pm$ 10 days. The ACF of the {\tt Everest}-reduced LC returns an amplitude consistent with zero, dominated by the leftover long-period trend that can be seen by-eye in the LC (Figure \ref{9978_K2LC}, third panel). The \emph{P$_{\rm quasi}$} values returned by successful fits for both targets are similar to each other and the strongest signals in all LC periodograms (Figure \ref{LC_pgrams}). The \emph{$\tau$} values are also all quite long relative to \emph{P$_{\rm quasi}$}. It is possible but unlikely that both stars coincidentally have similar rotation periods with very slow evolution of magnetic active regions.  Therefore, we cannot rule out the potential that the ACF signal is the result of remaining systematics for either target. We will rely on our analysis of the HARPS-N RVs in the next section to further search for periodic stellar activity signals.

\subsection{RV Stellar Activity Analysis}\label{rv_act_anal}

We first estimated values of $P_{\rm rot}$ for both targets using the \citet{Noyes84} correlation between $P_{\rm rot}$ and $R'_{HK}$, and applying a reddening correction of E(B-V) = 0.20 in the case of K2-79 \citep{green19}. This method yields a value of $P_{\rm rot, Noyes}$ $\approx$ 21.3d in the case of K2-79 ($\langle$S-index$\rangle$ = 0.159) and $P_{\rm rot, Noyes}$ $\approx$ 8.0d in the case of K2-222 ($\langle$S-index$\rangle$ = 0.160). We note that lower metallicity stars have higher $R'_{HK}$ values and therefore shorter estimated values of $P_{\rm rot}$, and vice versa, when using the \citet{Noyes84} correlation. Hence, for K2-79 $P_{\rm rot, Noyes}$ may be slightly overestimated ($\rm [Fe/H]$ = -0.05), and that of K2-222 may be underestimated ($\rm [Fe/H]$ = -0.25). We therefore consult other methods in further attempts to estimate $P_{\rm rot}$ from the RVs.

\begin{figure*}[]
\epsscale{1.2}
\plotone{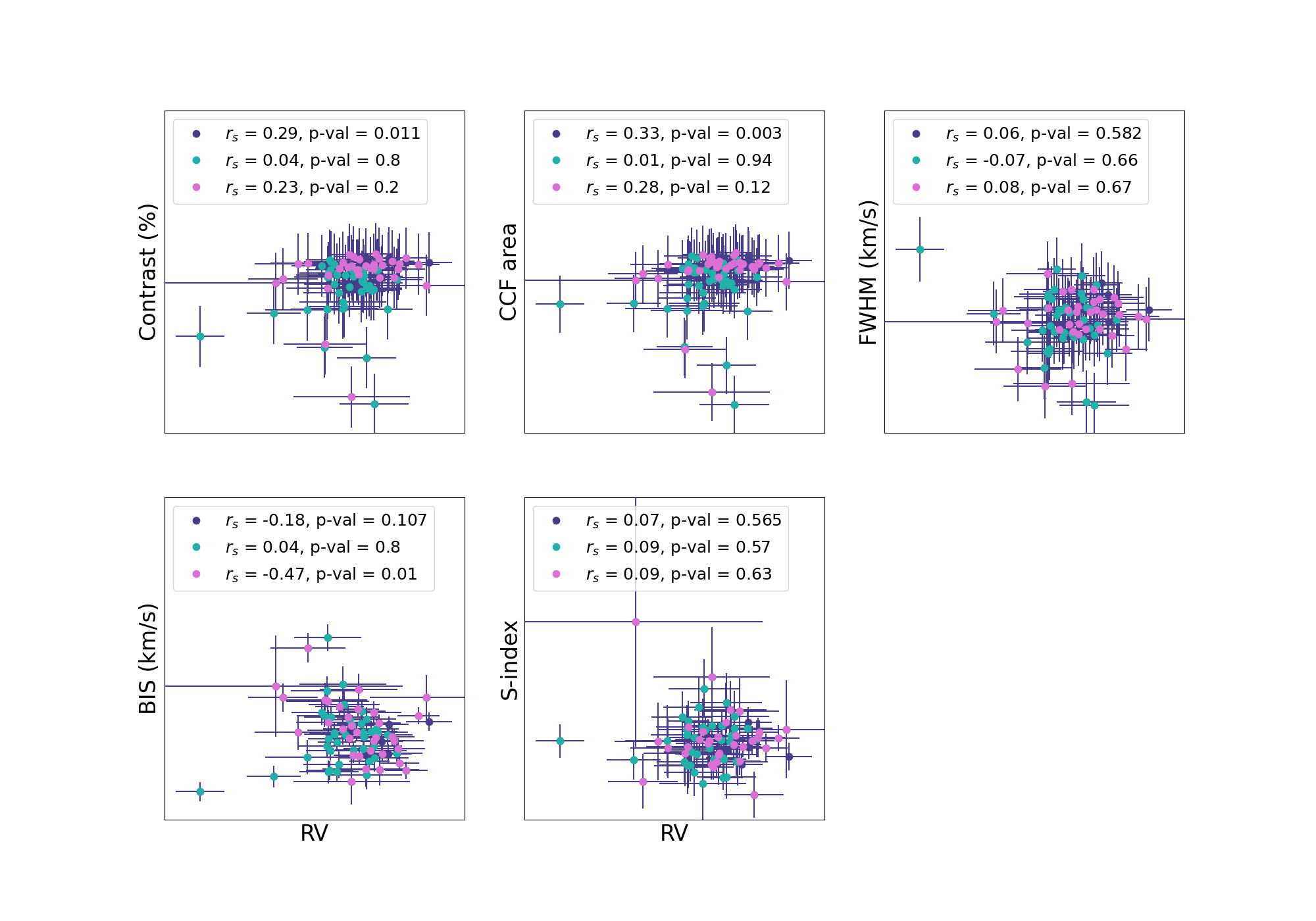}
\caption{Correlation plots and strengths (Spearman r coefficients, r$_{s}$) between the K2-79 RVs and various activity indicators (contrast, cross-correlation function area, full-width at half max., bisector inverse span, and S-index). All RVs and error-bars are plotted in dark purple, with season 1 RVs over-plotted in turquoise and season 2 RVs in magenta. P-value is a measure of how probable it is that the observed correlation is due to random chance.}\label{actcorr2237}
\end{figure*}

\begin{figure*}[]
\epsscale{1.2}
\plotone{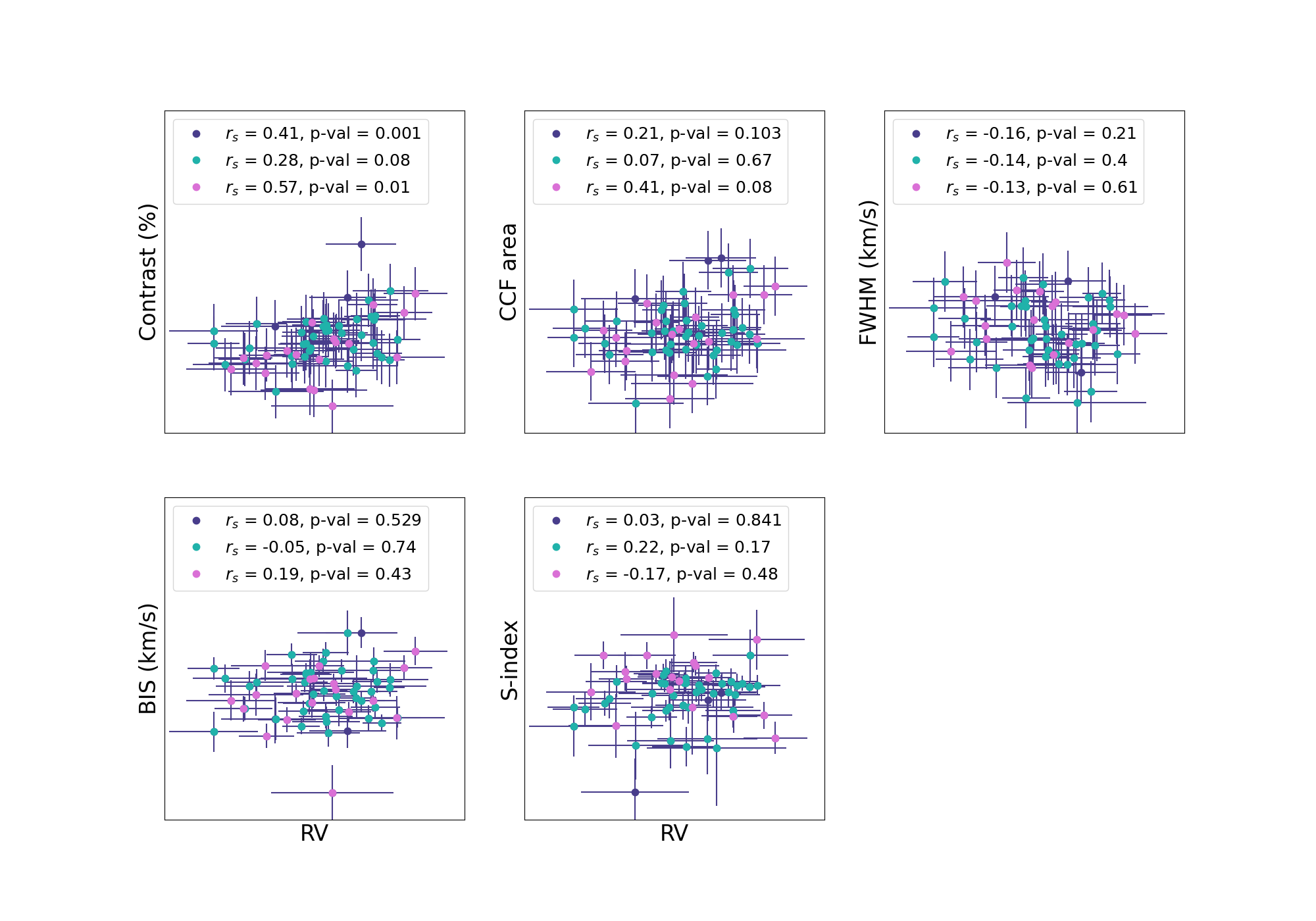}
\caption{Correlation plots and strengths (Spearman r coefficients, r$_{s}$) between the K2-222 RVs and various activity indicators (contrast, cross-correlation function area, full-width at half max., bisector inverse span, and S-index). All RVs and error-bars are plotted in dark purple, with season 1 RVs covered in turquoise and season 2 RVs covered in magenta. P-value is a measure of how probable it is that the observed correlation is due to random chance. }\label{actcorr9978}
\end{figure*}

To investigate the overall strength of stellar activity signals in the RVs, we calculated Spearman correlation coefficients (r$_{s}$) between the RVs and each of the measured activity indicators. In order to identify changes from season to season, we calculated three sets of correlation coefficients for each target: one for the full RV data set, one with just the first season (S1) RVs, and one with just the second season (S2) RVs. Other observing seasons contained too few points for a meaningful calculation. Figures \ref{actcorr2237} and \ref{actcorr9978} show correlation plots and associated r$_{s}$ values between the RVs and each of the activity indicators for K2-79 and K2-222, respectively. The p-value, a measure of how probable it is that the observed correlation would be produced by a null dataset, is also shown. To compare correlation strengths, we establish the following terminology corresponding to the absolute value of r$_{s}$:

\begin{itemize}
\item $<$ 0.20 = no correlation 
\item 0.20 - 0.39 = weak correlation
\item 0.40 - 0.59 = moderate correlation
\item $\geq$ 0.60 = strong correlation
\end{itemize}

In the case of K2-79, the full RV set has weak correlations with the contrast, CCF area, and BIS. The S1 RVs have no correlation with any activity indicators. The S2 RVs have weak correlations with the contrast and CCF area and a moderate anti-correlation with the BIS, where we see the strongest peak at 11.0d. In the case of K2-222, the full RV set has a moderate correlation with the contrast, where we see the strongest peak at 39.2d, and a weak correlation with the CCF area. The S1 RVs have weak correlations with the contrast and S-index. The S2 RVs have moderate correlations with the contrast and CCF area, where we see the strongest peak at 41.8d in both. In the cases of both targets, the S2 RVs show greater overall correlation with activity indicators than the S1 RVs, which we also see in the periodogram analysis.


\begin{figure*}[]
\epsscale{1.2}
\plotone{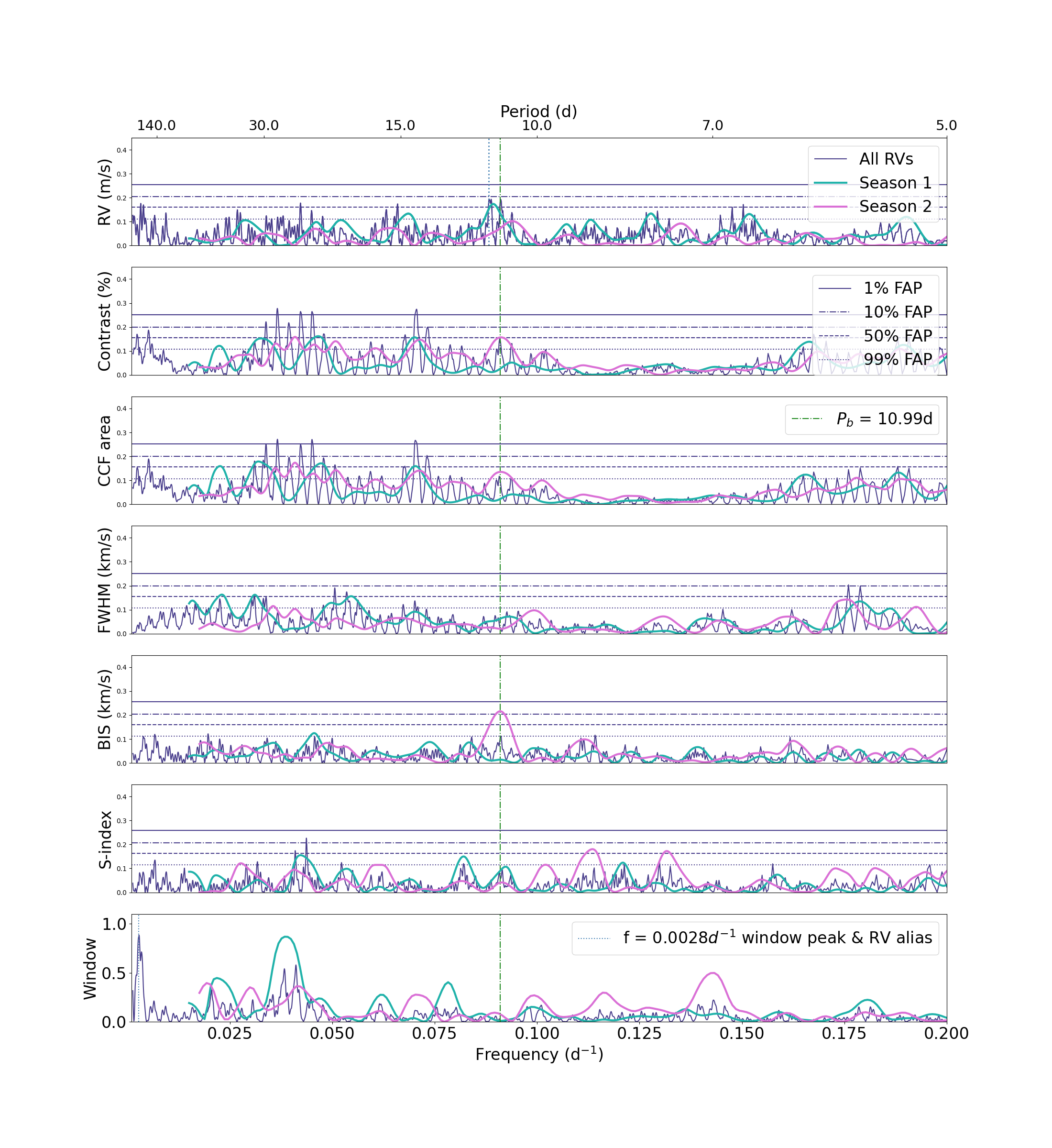}
\caption{K2-79 GLS periodograms of HARPS-N RVs and activity indicators. The full RV set shows weak correlations with the contrast, CCF area, and BIS. The S1 RVs show no correlation with any activity indicators. The S2 RVs show weak correlations with the contrast and CCF area, as well as a strong anti-correlation with the BIS. P$_{b}$ is the period of the known transiting exoplanet. The signal at P$_{b}$ could potentially be combined with an activity signal, based on the corresponding peaks with FAP $\approx 10\%$ and $\approx 50\%$ in the S2 BIS and contrast periodograms, respectively. At P$_{b}$, the seasonal RVs show a stronger signal in S1, but the seasonal activity indicators show stronger signals in S2, particularly in contrast and BIS.} \label{2237_RVpgram_freq}
\end{figure*}

\begin{figure*}[]
\epsscale{1.2}
\plotone{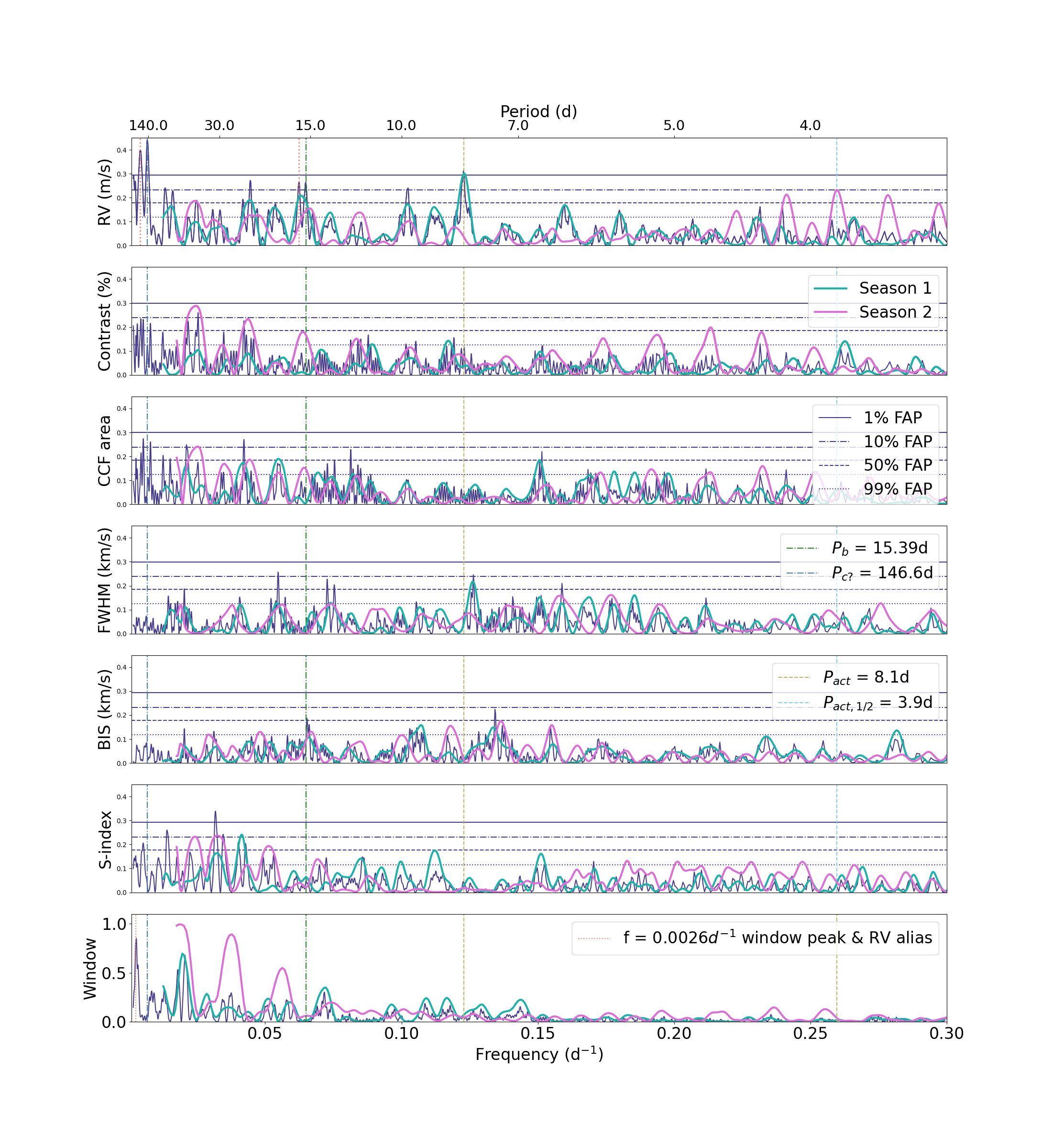}
\caption{K2-222 GLS periodograms of HARPS-N RVs and activity indicators. The full RV set shows a moderate correlation with the contrast and a weak correlation with the CCF area. The S1 RVs show weak correlations with the contrast and S-index. The S2 RVs show moderate correlations with the contrast and CCF area. P$_{b}$ is the period of the known transiting exoplanet. The signal at P$_{b}$ could potentially be combined with an activity signal, based on the corresponding peak with FAP $\approx 10\%$ in the contrast periodogram. At P$_{b}$, the seasonal RVs show a stronger signal in S1, but the seasonal activity indicators show stronger signals in S2, particularly contrast. P$_{c?}$ is the period of a potential second companion object, and P$_{act}$ with FAP $\approx 1\%$ is a signal we attribute to stellar activity because of the nearby peaks with FAP $\approx 10\%$ in the FWHM periodograms. The signal at P$_{act,1/2}$ with FAP $\approx 10\%$ may also be related to activity, as it occurs at approximately one-half of P$_{act}$ and less than 1d away from a peak with FAP $<$ 50\% in the contrast periodogram. The four prominent peaks surrounding the peak at P$_{act,1/2}$ are aliases with the largest peak in the S2 window function.} \label{9978_RVpgram_freq}
\end{figure*}

We calculate GLS periodograms for the RVs and activity indicators of K2-79 and K2-222 using three sets of data for each: The full dataset, the first season (S1) dataset, and the second season (S2) dataset. The strongest periods in stellar activity signals can change significantly from season to season due the sizes, locations, and distribution of stellar features evolving over time. Table \ref{Nobs} provides more detailed information about the full and seasonal data sets. We also calculate the window functions using the method described in Section \ref{lc_act_anal}. The resultant sets of periodograms for each target are shown in Figures \ref{2237_RVpgram_freq} and \ref{9978_RVpgram_freq}. In the periodograms we only show up to f = 0.20d$^{-1}$ because no statistically significant peaks occur beyond that frequency in the case of K2-79, and, in the case of K2-222, statistically significant peaks beyond that frequency are aliases of the 1d peak in the window function. 

We address statistically significant peaks in the periodograms, as well as those that occur at the known orbital periods of the transiting exoplanets. Although none of the peaks at the known orbital periods meet the traditional criteria for statistical significance (FAP $<$ 1$\%$), the probability of the peaks being false decreases beyond traditional FAP levels with the confirmed existence of exoplanets at those periods. 

For both K2-79 and K2-222, magnetic activity signals generally show a stronger potential presence in S2 than in S1. This is apparent both in the relative strengths of the S1 and S2 seasonal activity indicator periodograms and the S1 and S2 Spearman correlation coefficients, especially considering the much lower number of data points in S2. For each target, the strongest power in the S2 activity indicator periodograms is greater than or comparable to the strongest in S1, in spite of far fewer points in the S2 data set (Figures \ref{2237_RVpgram_freq} and \ref{9978_RVpgram_freq}). In the case of K2-79, none of the S1 activity indicators correlate with the RVs, while three of the S2 activity indicators show at least weak correlation (Figure \ref{actcorr2237}). In the case of K2-222, two activity indicators correlate with the RVs in both S1 and S2. There is a correlation between the S1 S-index values and RVs (r$_{s}$ = 0.22) where no correlation exists in S2, but it is much weaker than the correlations between contrast and RVs for both S1 (r$_{s}$ = 0.28) and S2 (r$_{s}$ = 0.57), where the S2 Spearman coefficient is significantly higher (Figure \ref{actcorr9978}). 

For both targets, at $P_{\rm b}$ specifically, seasonal periodograms also show more potential for activity interference in S2 than in S1. Even though the K2-79 and K2-222 seasonal \emph{RV} periodograms reveal stronger signals near $P_{\rm b}$ in S1, the seasonal \emph{activity indicator} periodograms show stronger signals at $P_{\rm b}$ in S2. In the case of K2-79, the S1 RV peak near $P_{\rm b}$ has a FAP $\approx 10\%$, while the S2 RV peak has no statistical significance (FAP $< 99\%$). Yet, near $P_{\rm b}$, the S1 activity indicators contain no peaks of significance, while the S2 BIS contains a peak with FAP $< 10\%$, and the S2 contrast contains a peak with FAP $\approx 50\%$. In the case of K2-222, the S1 RV peak near $P_{\rm b}$ is far stronger (FAP $< 10\%$) than the S2 peak (FAP $> 50\%$), which is expected given the relative sizes of the seasonal data sets. However, near $P_{\rm b}$, the S1 activity indicators again contain no peaks of significance, while the S2 contrast contains a peak with FAP $\approx 50\%$. 

In the case of K2-79 (Figure \ref{2237_RVpgram_freq}), no statistically significant peaks occur in the full or either of the seasonal RV periodograms. The two strongest peaks occur at the known orbital period of the planet (P$_{b}$ = 10.99d; $f = 0.09099d^{-1}$) and at $f = 0.0882d^{-1}$, an alias of P$_{b}$ with the strongest peak in the full window function ($f_{w}$ = 0.0028$d^{-1}$). The signal at P$_{b}$ could potentially be combined with an activity signal, based on the corresponding peaks with FAP $\approx 10\%$ and $\approx 50\%$ in the S2 BIS and contrast periodograms, respectively, and its location less than one day from the first harmonics of the estimated $P_{\rm rot, max}$ and $P_{\rm rot, Noyes}$.

In the case of K2-222 (Figure \ref{9978_RVpgram_freq}), the strongest peak in the full RV periodogram occurs at $f = 0.007d^{-1}$ (P = 146.6d), notably not at the period of the known exoplanet. An alias of this signal with the strongest peak in the full window function ($f = 0.0026d^{-1}$) also occurs at $f = 0.0042d^{-1}$. The peak at 146.6d does not appear to be related to the window function or activity indicator periodograms; peaks at nearby frequencies do occur in the activity indicators, but not at the exact same frequency, and the signal is much more significant in the RVs. The 146.6d signal remains the most significant peak (FAP $<$ 1\%) when we remove the last three solitary observations and recalculate the periodogram. We therefore consider this signal as a potential second object (P$_{c?}$).  

The periodogram of the full K2-222 RV set also shows peaks at the period of the known transiting exoplanet (P$_{b}$ = 15.39d; f = 0.06498$d^{-1}$) and at f = 0.0624, an alias of the exoplanet signal with the strongest peak in the full window function ($f_{w}$ = 0.0026$d^{-1}$). The signal at P$_{b}$ could potentially be combined with an activity signal, based on the corresponding peak with FAP $\approx 10\%$ in the contrast periodogram. We analyze an additional statistically significant peak at f = 0.1229$d^{-1}$ (P$_{act}$ = 8.137d) that occurs in both the full and S1 periodograms. We believe the signal is most likely a result of stellar activity because of a nearby peak with FAP $\approx 10\%$ in the FWHM periodograms. The signal at P$_{act,1/2}$ with FAP $\approx 10\%$ may also be related to activity, as it occurs at approximately one-half of P$_{act}$ and less than 1d away from a peak with FAP $<$ 50\% in the contrast periodogram. The four prominent peaks surrounding the peak at P$_{act,1/2}$ are aliases with the largest peak in the S2 window function.

As discussed above, for both K2-79 and K2-222, the S2 RVs suffer more potential interference from stellar activity at $P_{\rm b}$ than the S1 RVs. Since each target's S1 and S2 observations were not taken in the same seasons as the other's, we can rule out any seasonal instrumental effect being the cause of more interference in S2. For K2-79 and K2-222, the small number of observations in S2 will already make for a less \emph{precise} mass determination, but constructive or destructive interference with stellar activity in the S2 RVs could also affect the \emph{accuracy} of the exoplanet mass measured from the full RV set. Therefore, in addition to full RV fits, we perform fits to the S1 and S2 RV sub-sets to confirm agreement between exoplanet masses estimated using the full result and at least the S1 result. 

\section{Simultaneous Radial Velocity/Transit Exoplanet Fit} \label{simult_fit}

Attempts to fit the K2-79 and K2-222 LCs and RVs separately revealed that for both targets, neither of the data sets is sufficient to constrain the orbital eccentricity of the transiting exoplanet. We therefore perform a simultaneous fit to the RVs and flattened LCs (top panels Figures \ref{2237_RVscat} and \ref{9978_RVscat}, bottom panels of Figures \ref{2237_K2LC} and \ref{9978_K2LC}) using the publicly available Differential Evaluation Markov-Chain Monte-Carlo (MCMC) software, {\tt EXOFAST v2} \citep{eastman19}. We run our {\tt EXOFAST v2} fits with \emph{maxsteps = 50,000} and \emph{nthin = 30}, running for a maximum of 50,000 recorded steps while recording the 30th step of each walker. The global model used in {\tt EXOFAST v2} includes spectral energy distribution and integrated MIST stellar evolutionary models informed by the magnitudes in Table \ref{starpars} to constrain stellar parameters. In the case of both targets, we utilize isochrones from \citet{MIST} to derive a physical star (mass, radius, luminosity, age) based on $T_{\rm eff}$, $\rm [Fe/H]$, and  $\logg$.
Table \ref{exofit_startrans} shows the median and 1$\sigma$ stellar values returned for K2-79 and K2-222.

For the transit fits, we set the \emph{longcadence} flag to account for the 29.4 min cadence of the K2 data, and we use the default quadratic limb darkening law, based on tables reported in \cite{claret11}. The transit model includes parameters corresponding to each transiting exoplanet's orbital period (P$_{\rm orb}$), time of central transit ($T_{\rm c}$), orbital inclination (i), orbital eccentricity (e), orbital argument of periastron ($\omega$), and radius relative to its star ($R_P/R_*$). The RV model is a Keplerian consisting of the following parameters: RV semi-amplitude (K), e, P$_{\rm orb}$, and $\omega$. 

The transit fits place strong constraints on P$_{\rm orb}$ and $T_{\rm c}$. Fit alongside with stellar density, they also drastically reduce the degenerate e-$\omega$ parameter space. Pairing the transit and RV fits therefore allows us to break the degeneracy between these two parameters and achieve better estimates of e and $\omega$ than were possible with the RVs or transits alone. 

For each target, we perform three fits assuming a single planet: one using all available RVs, another with just the S1 RVs, and a third with just the S2 RVs. In the case of K2-222, we perform a second set of these three fits for a 2-planet scenario to investigate the nature of the strongest RV periodogram signal at 146.6d. We fix the eccentricity of the potential second exoplanet to zero, taking into account the exoplanet's relatively long period and gaps in its phase coverage. 

Table \ref{exofast_priors} summarizes the probability distributions of priors applied to the {\tt EXOFAST v2} fits for both targets. Priors on stellar parameters were based on the combined result from the Stellar Parameter Classification tool and ARES+MOOG analyses of the HARPS-N spectra (see Section \ref{starchar} and Table \ref{starpars}). In the case of the seasonal 2-planet fits to K2-222 data, we fix the parameters associated with the potential second exoplanet to the median values returned by the full 2-planet fit. This allows us to confirm agreement between the full and seasonal RV sets of the parameters associated with K2-222b, despite the fact that the seasonal baselines are too short to reliably fit the 146.6d signal of a potential second companion object. 

Table \ref{exofit_startrans} shows the MCMC posterior median values and 1$\sigma$ uncertainties of the stellar and exoplanet transit parameters for each target fitted with {\tt EXOFAST v2}. We also list several other parameters describing the individual transits (depth, duration, impact parameter) and full LCs (baseline flux, variance of transit model residuals). The estimated equilibrium temperature (T$_{\rm eq}$) returned by {\tt EXOFAST v2} follows Equation 1 of \citet{Hansen2007}, $T_{\rm eq} = \teff\sqrt{R_{*}/(2\emph{a})}$, where $\emph{a}$ is orbital semi-major axis and no albedo and perfect energy redistribution are assumed. The stellar values returned by the {\tt EXOFAST v2} fits agree with those adopted from our stellar analysis within 1$\sigma$, with the exception of K2-222's stellar radius, which sees a slightly higher discrepancy but agrees well within 1.5$\sigma$. 

Tables \ref{exofit_2237_tab} and \ref{exofit_9978_tab} show the posterior median values and 1$\sigma$ uncertainties of all other parameters from the {\tt EXOFAST v2} fits to K2-79 and K2-222, respectively. In the cases of both targets, the obtained K values for K2-79b and K2-222b from the full RVs and seasonal fits agree within 1$\sigma$. 
In the case of K2-222, the 1-planet and 2-planet fits also return values consistent within 1$\sigma$, but median value for the semi-amplitude of the transiting planet, K$_{\rm b}$, is notably smaller in the 2-planet model. 

We find evidence in support of a second companion candidate by comparing the robustness of the 1-planet and 2-planet models using their Bayesian information criterion (BIC) and Akaike information criterion (AIC) values. The BIC value is lower for the 1-planet model full fit, but the AIC value is lower for the 2-planet model by a much larger margin. The AIC and BIC of the 2-planet model are also penalized more due to the higher number of parameters fit than in the 1-planet model. Although the AIC and BIC values provide tentative support in favor of a 2-planet model, we adopt the 1-planet solution for now, considering the significant phase gaps in the coverage of the 147.5d period returned by our fit.

\begin{table*}
   \centering
    \caption{Prior probability distributions applied to parameters modelled in our {\tt EXOFAST v2} transit fits. Stellar priors are based on combined results from Stellar Parameter Classification tool and ARES+MOOG analyses of the HARPS-N spectra. The two numbers in each prior are the mean and standard deviation of the Gaussian distribution, respectively.} \label{exofast_priors}
    \begin{tabular}{lcc}
        \hline
        Target & K2-79 & K2-222 \\
        \hline
         T$_{\rm eff}$ [K] &  Gaussian(5897, 58.5) & Gaussian(5942, 57.5) \\
         Surface gravity, $\logg$ & Gaussian(4.25, 0.15) & Gaussian(4.22, 0.16)  \\
         Metallicity, $\rm [Fe/H]$ & Gaussian (0.035, 0.06) & Gaussian (-0.315, 0.06) \\
        \hline
    \end{tabular}
    
\end{table*}

\begin{table*} 
\caption{K2-79 and K2-222 stellar and transit-related planet parameters based on median values and 68\% confidence interval returned by {\tt EXOFAST v2} simultaneous RV and LC fits. Added variance describes remaining scatter of the data after removal of the transit model.} \label{exofit_startrans}
         \small
         \centering
\begin{tabular}{lccr}
\hline
~~~Symbol & Parameter and Units & K2-79 & K2-222~~~ \\
\hline
\\
Stellar Parameters & & & \\
~~~~$M_*$\dotfill & Mass ($\msun$)\dotfill &$1.066^{+0.057}_{-0.070}$ & $0.989^{+0.070}_{-0.065}$ \\ 
~~~~$R_*$\dotfill &Radius ($\rsun$)\dotfill &$1.265^{+0.041}_{-0.027}$ & 1.115 $\pm$ 0.029 \\
~~~~$L_*$\dotfill &Luminosity ($\lsun$)\dotfill & 1.76 $\pm$ 0.10 & $1.452 \pm 0.077$ \\
~~~~$Age$\dotfill &Age (Gyr)\dotfill &$6.6^{+2.9}_{-2.6}$ & $6.3^{+3.4}_{-3.0}$ \\
~~~~$T_{\rm eff}$\dotfill & Effective Temperature (K)\dotfill &$5902^{+86}_{-59}$ & $6000^{+67}_{-69}$ \\
~~~~$\log{g}$\dotfill &Surface gravity (cgs)\dotfill & $4.257^{+0.038}_{-0.043}$ & 4.339 $\pm$ 0.042 \\
~~~~$[Fe/H]$\dotfill & Metallicity \dotfill &$0.053^{+0.048}_{-0.036}$ & -0.143$^{+0.071}_{-0.14}$ \\
\\
Exoplanet Parameters: & & b & b \\
~~~~$P$\dotfill &Period (days)\dotfill &$10.99470^{+0.00031}_{-0.00047}$ & $15.38857 \pm 0.00088$ \\
~~~~$T_C$\dotfill &Time of Transit (BJD - 2.4E6)\dotfill & 57103.22750$^{+0.00076}_{-0.00084}$ & $57399.0595 \pm 0.0016$\\
~~~~$i$\dotfill &Inclination (Degrees)\dotfill & 88.44 $\pm$ 0.44 & $89.12^{+0.55}_{-0.41}$ \\
~~~~$R_P/R_*$\dotfill &Radius of planet in stellar radii \dotfill & 0.02948$^{+0.00094}_{-0.00037}$ & $0.01928^{+0.00040}_{-0.00035}$ \\
\\
Transit Parameters:&&& \\
~~~~$F_0$\dotfill & Baseline flux \dotfill & 0.9999949 $\pm$ 0.0000095 & $1.0000015 \pm 0.0000060$\\
~~~~$\delta$\dotfill &Transit depth (fraction)\dotfill &$0.000869^{+0.000057}_{-0.000022}$ & $0.000372^{+0.000016}_{-0.000013}$\\
~~~~$T_{14}$\dotfill &Total transit duration (days)\dotfill & $0.1894^{+0.0026}_{-0.0029}$ & $0.1836^{+0.0028}_{-0.0025}$\\
~~~~$b$\dotfill &Transit Impact parameter \dotfill &$0.46^{+0.15}_{-0.16}$ & $0.31^{+0.19}_{-0.20}$\\
~~~~$\sigma^{2}$\dotfill &Added Variance \dotfill &$2.28x10^{-9} \pm 9.6x10^{-10}$ & $1.85x10^{-9} \pm 3.7x10^{-10}$ \\
\hline
\end{tabular}
\end{table*}

\begin{table*} 
\caption{K2-79 radial velocity parameters based on median values and 1$\sigma$ confidence interval returned by {\tt EXOFAST v2}  simultaneous RV and LC fits.} \label{exofit_2237_tab}
         \small
         \centering
\begin{tabular}{lcccr}
\hline
~~~Symbol & Parameter and Units & All & S1 & S2 \\
\hline
\\
Exoplanet Parameters: &&&& \\ 
~~~~$K_{b}$\dotfill &RV semi-amplitude (m/s)\dotfill &$3.28^{+1.0}_{-0.78}$ & $3.2^{+1.1}_{-1.3}$ & $3.0^{+1.6}_{-1.4}$ \\
~~~~$e_{b}$\dotfill &Eccentricity \dotfill &$0.082^{+0.088}_{-0.056}$ & 0.133$^{+0.14}_{-0.095}$ & 0.119$^{+0.12}_{-0.079}$ \\ 
~~~~$\omega_{b}$\dotfill &Argument of Periastron (Degrees)\dotfill & $-140^{+150}_{-130}$ & $5^{+73}_{-50}$ & $160 \pm 110$  \\
\\
HARPS-N Parameters: &&&& \\
~~~~$\gamma_{\rm rel}$\dotfill &Relative RV Offset (m/s)\dotfill &$-0.06^{+0.67}_{-0.73}$ & $-2.10^{+1.2}_{-0.99}$ & $1.1 \pm 1.2$ \\
~~~~$\sigma_J$\dotfill &RV Jitter (m/s)\dotfill &$3.55 \pm 0.67$ & $3.7 \pm 1.1$ & $2.4^{+1.4}_{-2.4}$\\
\hline
\end{tabular}
\end{table*}

\begin{table*}
\caption{K2-222 radial velocity parameters based on median values and 68\% confidence interval returned by {\tt EXOFAST v2}  simultaneous RV and LC fits.} \label{exofit_9978_tab}
\hspace{-1.5cm}
\begin{tabular}{lccccccr} 
\hline
~~~Symbol & Parameter and Units & All/1p & All/2p & S1/1p & S1/2p & S2/1p & S2/2p \\
\hline
\\
Exoplanet Parameters:&&&&&&& \\
~~~~$K_{b}$\dotfill &RV semi-amplitude (m/s)\dotfill &$2.18^{+0.48}_{-0.50}$ & $1.65^{+0.50}_{-0.51}$ & $2.26^{+0.69}_{-0.62}$ & $1.76^{+0.52}_{-0.45}$ & $2.52^{+1.1}_{-0.92}$  & $1.95^{+0.73}_{-0.76}$\\
~~~~$e_{b}$\dotfill &Eccentricity \dotfill &$0.188^{+0.11}_{-0.097}$ & 0.23$^{+0.11}_{-0.12}$ & 0.30$^{+0.13}_{-0.19}$ & 0.21$^{+0.13}_{-0.12}$ & 0.17$^{+0.14}_{-0.11}$  & 0.144$^{+0.12}_{-0.087}$ \\
~~~~$\omega_{b}$\dotfill & Argument of Periastron (Degrees)\dotfill & $151^{+30}_{-44}$ & $164^{+29}_{-38}$ & $163^{+52}_{-64}$ & $157^{+27}_{-52}$ & $162^{+52}_{-81}$   & $145^{+59}_{-87}$ \\
~~~~$K_{c}$\dotfill & RV semi-amplitude (m/s)\dotfill & - & $2.52^{+0.52}_{-0.51}$ & - & $\equiv$ 2.52 & - & $\equiv$ 2.52\\
~~~~$P_{c}$\dotfill & Period (days)\dotfill & - & $147.5 \pm 3.3$ & - & $\equiv$ 147.5 & - & $\equiv$ 147.5\\
~~~~$e_{c}$\dotfill &Eccentricity \dotfill & - & $\equiv$ 0 & - & $\equiv$ 0 & - & $\equiv$ 0\\
~~~~$\omega_{c}$\dotfill &Argument of Periastron (Degrees)\dotfill & - & $\equiv$ 0 & - & $\equiv$ 0 & -  & $\equiv$ 0 \\
\\
HARPS-N Parameters:&&&&&&& \\
~~~~$\gamma_{\rm rel}$\dotfill & Relative RV Offset (m/s)\dotfill &$0.04^{+0.35}_{-0.37}$ & $-0.10 \pm 0.31$ & $0.32^{+0.43}_{-0.48}$  & $-0.01^{+0.40}_{-0.35}$ & $-0.60^{+0.73}_{-0.72}$ & $-0.21^{+0.52}_{-0.51}$\\
~~~~$\sigma_J$\dotfill &RV Jitter (m/s)\dotfill &$1.98^{+0.36}_{-0.31}$ & $1.26^{+0.38}_{-0.40}$ & $2.13^{+0.46}_{-0.48}$ & $1.46^{+0.59}_{-0.46}$ & $2.26^{+0.85}_{-0.76}$ & $0.89^{+0.96}_{-0.89}$\\
\\
Model Comparison: &&&&&&& \\
~~~~N$_{RV}$\dotfill & Number of RV observations \dotfill & 63 & 63 & 41 & 41 & 19 & 19 \\
~~~~AIC\dotfill & Akaike information criterion \dotfill & -1454.3836 & -1468.3867 & -1538.5536 & -1567.6584 & -1665.0455 & -1667.2941\\
~~~~$\Delta$AIC\dotfill & AIC difference from adopted \dotfill & 0 & -14.0031 & -84.1700 & -113.2748 & -210.6619 & -212.9105 \\
~~~~BIC\dotfill & Bayesian information criterion \dotfill & -1377.2251 & -1375.1535 & -1464.4513 & -1481.2057 & -1594.4461 & -1584.9282\\
~~~~$\Delta$BIC\dotfill & BIC difference from adopted \dotfill & 0 & +2.0716 & -87.2662 & -103.9806 & -217.2209 & -207.7031 \\
\hline
\end{tabular}
\end{table*}

\begin{figure*}[]
\epsscale{1.2}
\plotone{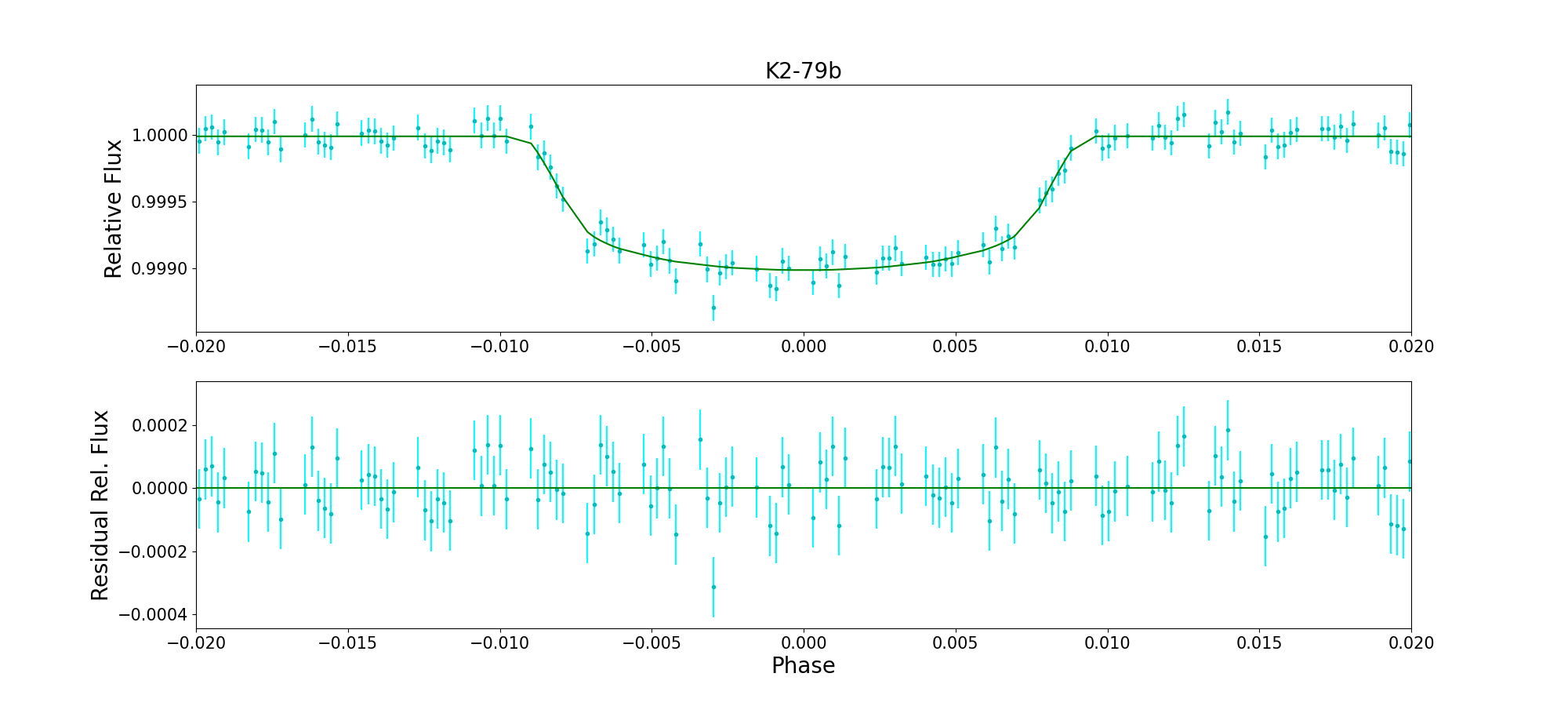}
\caption{Phase-folded K2-79 LC (cyan points) with the fit posterior transit model over-plotted (green line).}\label{2237_exofit}
\end{figure*}

\begin{figure*}[]
\epsscale{1.2}
\plotone{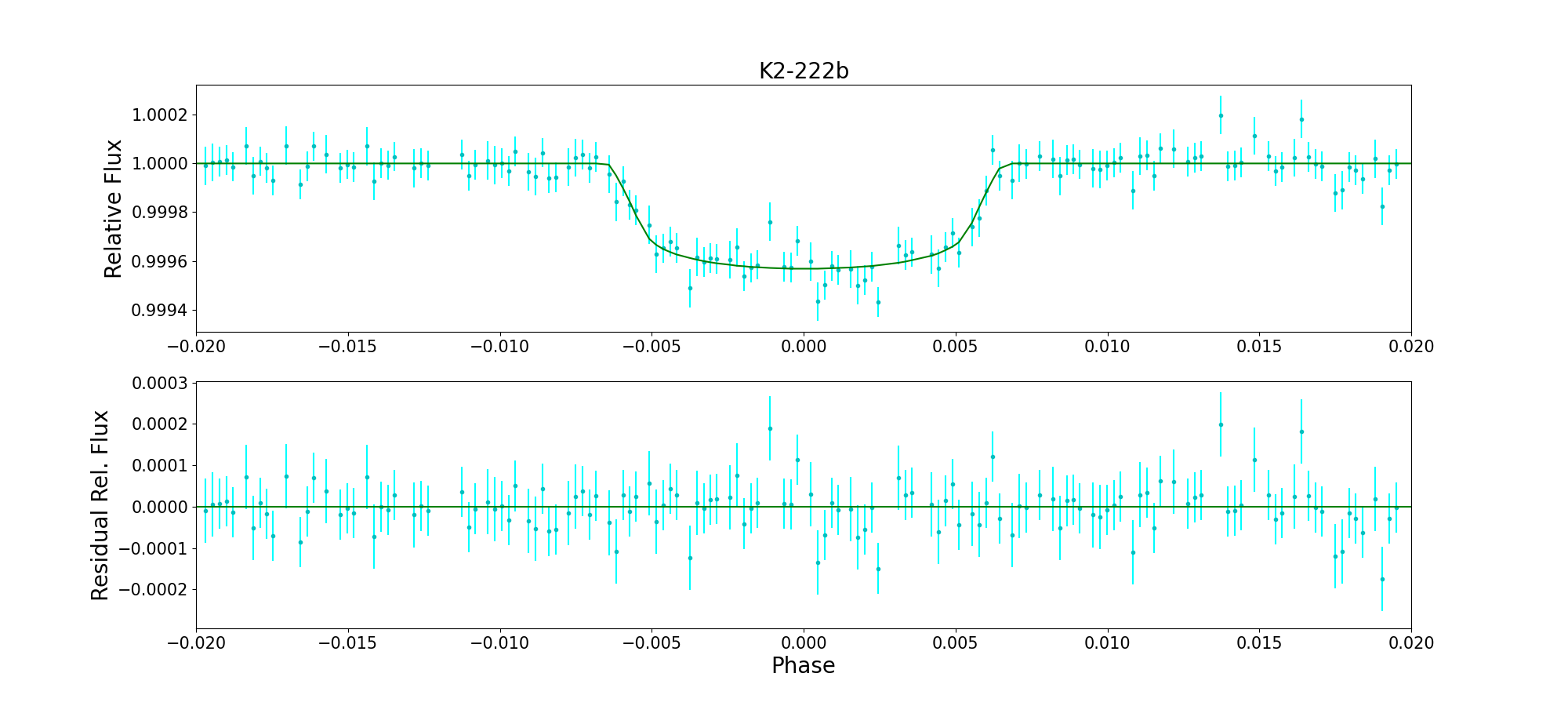}
\caption{Phase-folded K2-222 LC (cyan points) with the fit posterior model over-plotted (green line).}\label{9978_exofit}
\end{figure*}

\begin{figure*}[]
\epsscale{1.2}
\plotone{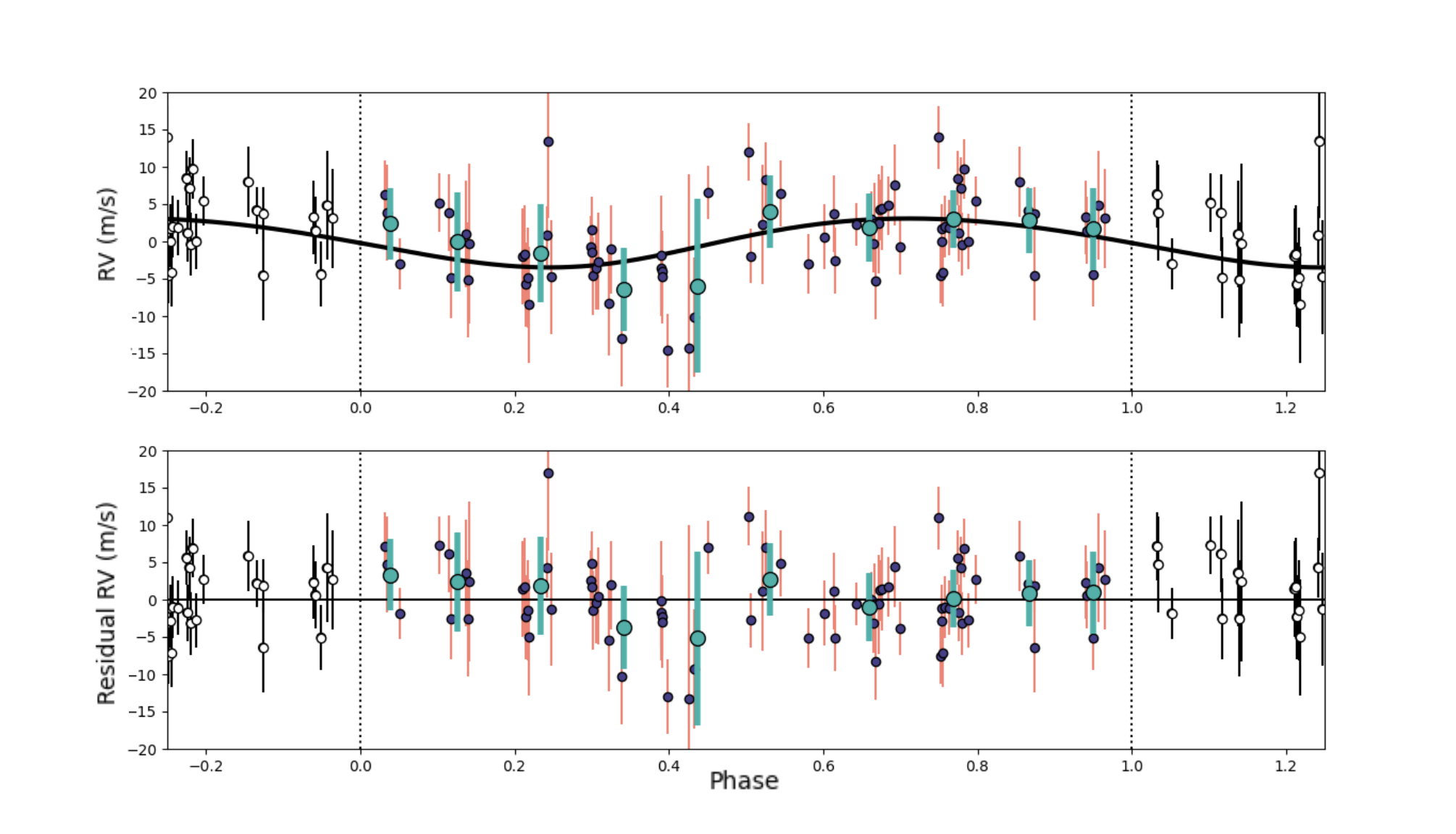}
\caption{Phase-folded K2-79b RV curve (purple points, orange error bar) with the fit posterior model over-plotted (black line). The binned RVs (teal), spaced equally in phase (binsize = 1/N$_{bins}$) and averaged over phase and RV, were not used in any fits.} \label{2237_RVfit}
\end{figure*}

\begin{figure*}[]
\epsscale{1.2}
\plotone{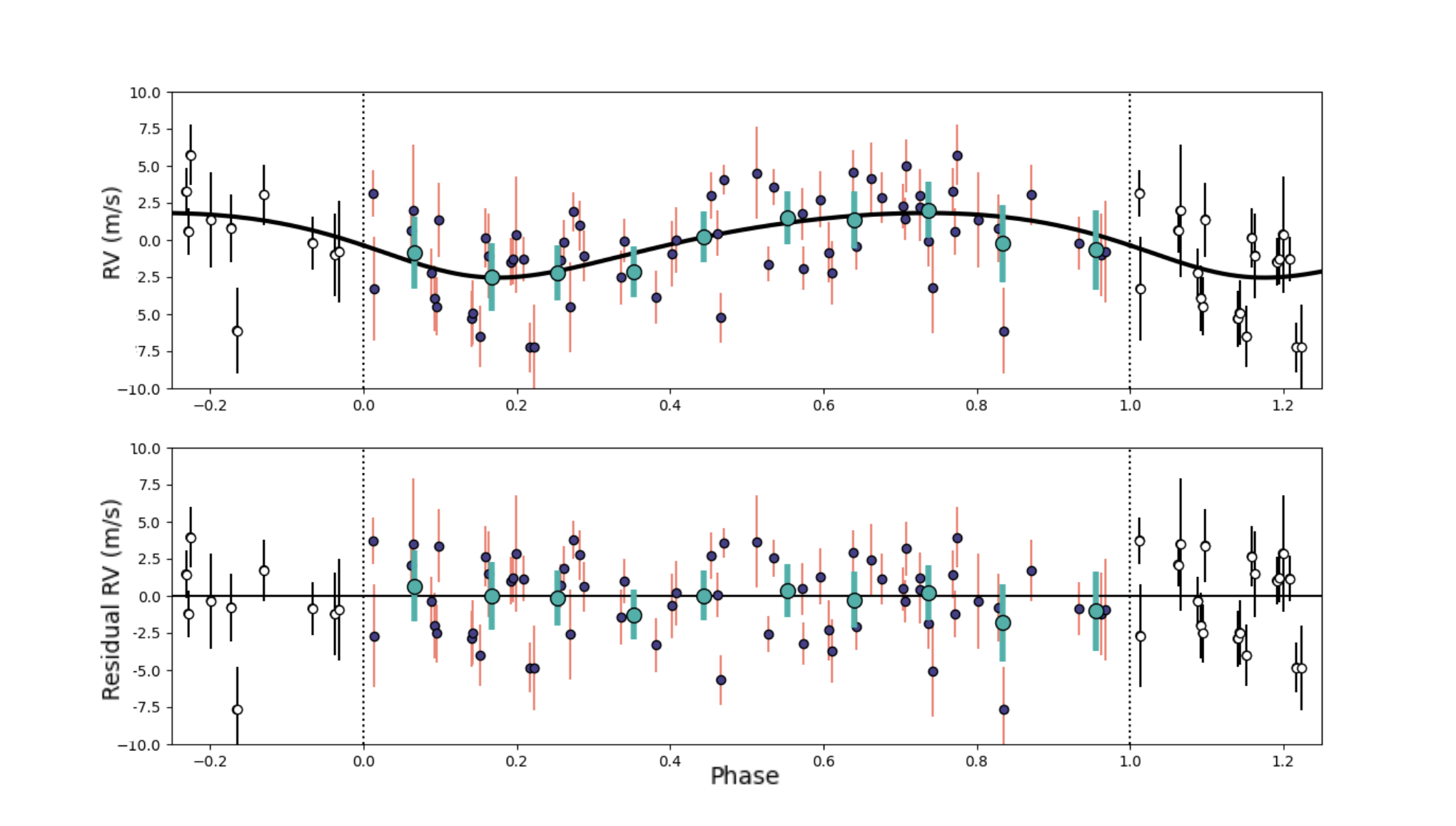}
\caption{Phase-folded K2-222b RV curve (purple points, orange error bar) with the 1-planet fit posterior model over-plotted (black line). The binned RVs (teal), spaced equally in phase and averaged over phase and RV, were not used in any fits.} \label{9978_RVfit}
\end{figure*}

\begin{figure*}[]
\epsscale{1.2}
\plotone{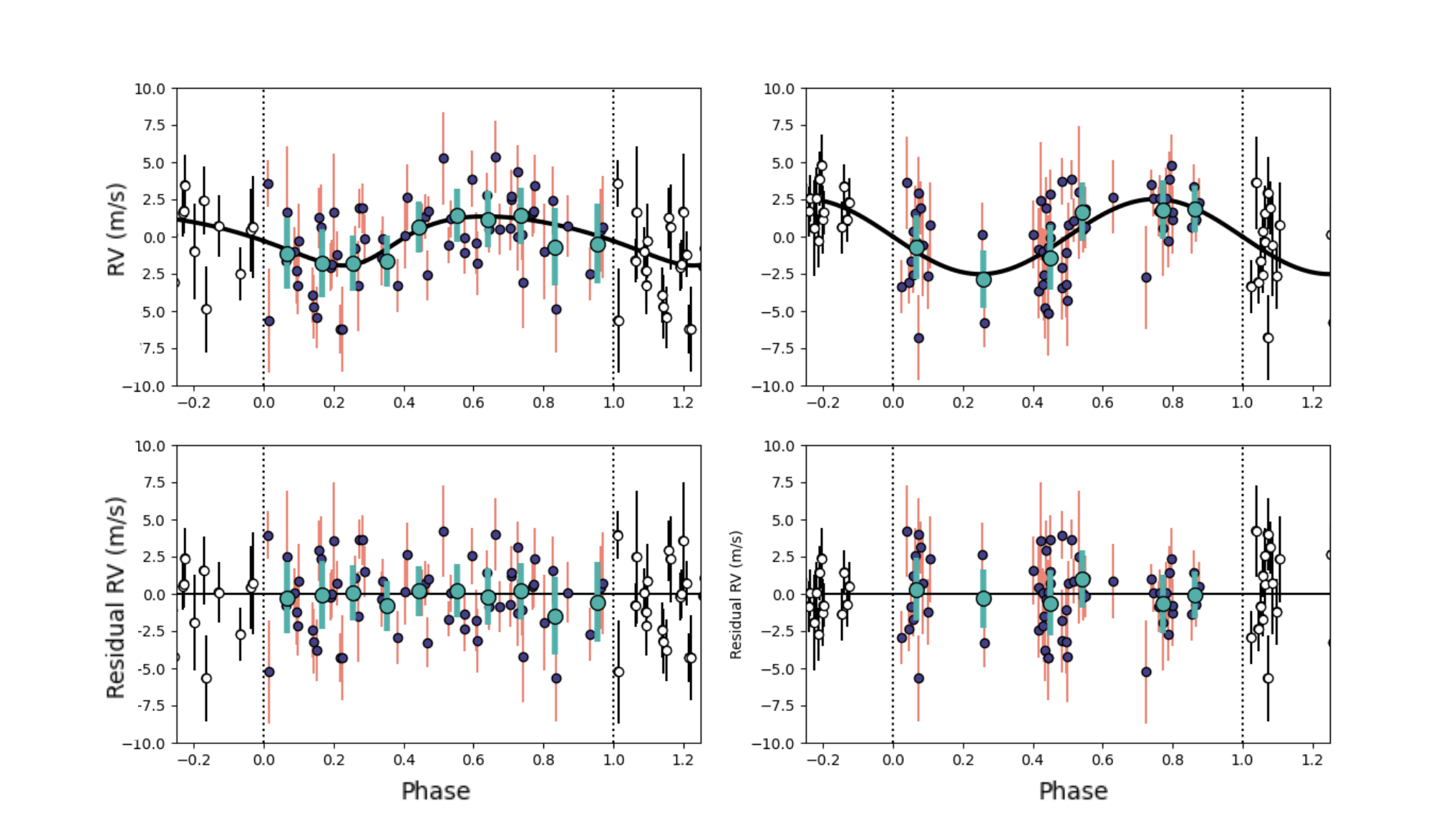}
\caption{Phase folded K2-222b (left) and candidate second companion (right) RVs (purple points, orange error bar) with the 2-planet fit posterior model over-plotted (black line). The binned RVs (teal), spaced equally in phase and averaged over phase and RV, were not used in any fits.} \label{9978_2plan_RVfit}
\end{figure*}

Figures \ref{2237_exofit} and \ref{9978_exofit} show the phase-folded LCs of K2-79b and K2-222b with the best-fit transit models over-plotted. Figures \ref{2237_RVfit} and \ref{9978_RVfit} show the phase-folded RV curves of each transiting exoplanet with the best fit 1-planet Keplerian model over-plotted. Figure \ref{9978_2plan_RVfit} shows the phase-folded RV curves of K2-222b and the candidate K2-222c with their best fit 2-planet Keplerian model over-plotted.  

Finally, in addition to the simultaneous RV and LC fits with EXOFASTv2, we also fitted the RV data with the {\tt RadVel} software \citep{RadVel} to see if we could improve the RV solutions using Gaussian Processes (GP) regression to model the stellar activity \citep{haywood14}. We detail these fits in the next section. 

\section{Radial Velocity Fit with Gaussian Processes Regression}\label{radvel_fit}

The following RV fits including GPs were not included in the calculation of any final mass results. However, in this section we detail our analysis and results for completeness.

In addition to the simultaneous RV and LC fits, we also use the {\tt RadVel} software \citep{RadVel} to perform MCMC fits on just the HARPS-N RVs, but this time including a quasi-periodic Gaussian Processes (GP) regression to model a stellar activity signal in addition to the exoplanet Keplerian \citep{haywood14}. We once again perform fits on the full and seasonal RV sets, and in the case of K2-222, with both 1-planet and 2-planet models.

The {\tt RadVel} model fits for the same Keplerian and instrumental parameters fit by {\tt EXOFAST v2}. In the 2-planet models we once again fix the eccentricity of the potential second exoplanet and, in the seasonal fits, fix all of its associated parameters to the values returned by the full fit. We place priors on $P_{\text{b}}$, $T_{\text{C,b}}$, and $e_{\rm b}$ for both targets based on posteriors from our {\tt EXOFAST v2} fits. We set Jeffreys priors on $K_{\rm b}$ values with an upper limit of 10m/s to explore parameter space around the $\approx$ 6m/s standard deviation of both RV sets. We place a Gaussian prior on $\sigma_{\rm HARPS-N}$, set equal to the distribution of RV errors ($\langle RVerr \rangle$, $\sigma_{RVerr}$). Table \ref{radvel_priors} summarizes the priors applied to exoplanet parameters in our {\tt RadVel} fits. 

The GP-generated quasi-periodic stellar activity model populates a covariance matrix with the function:

\begin{equation} 
\noindent
 k(t, t') = A^2  \exp\bigg[-\frac{(t-t')^2}{\tau^2}
 - \frac{sin^2\left(\frac{\pi (t-t')}{P_{\rm{quasi}}}\right)}{2\eta^2}\bigg],
\end{equation} \label{GP_eq}

where \emph{k(t,t')} is the correlation weight between observations taken at times t and t'. \emph{A} is the mean amplitude of the activity signal, \emph{$\tau$} is a hyper-parameter related to the evolution timescale of activity features, \emph{$P_{\rm quasi}$} is related to the stellar rotation period, and \emph{$\eta$} describes the level of high-frequency variation expected within a single stellar rotation and is related to the average distribution of activity features on the surface of the star.

We apply Jeffreys priors on the GP stellar amplitudes with the same limits and motivation applied to the $K_{b}$ priors, and also place a Jeffreys prior on \emph{$\tau$} investigating a wide range of values between 2d and 100d. As mentioned above, \emph{$\eta$} is physically related to the average distribution of magnetic active regions on the stellar surface. Models by \citet{jeffers} demonstrated that even highly complex activity distributions will average to just two to three large active regions in a given rotation. We therefore assign a prior distribution of \emph{$\eta$} $= 0.25 \pm 0.025$, which allows for two to three local minima or maxima per rotation \citep{haywood14, Grunblatt, LMH}.

For each target, we ran fits with various priors on \emph{$P_{\rm{quasi}}$}, first with a Jeffreys prior of 2-40d for each. Based on the posterior distributions returned by those fits, we ran additional fits with Jeffreys priors of 10-40d and 2-10d for K2-79 and K2-222, respectively. We also used several Gaussian priors centered around various $P_{\rm rot}$ estimates with a 20\% standard deviation. For K2-79 we ran fits with Gaussian priors on \emph{$P_{\rm{quasi}}$} centered around 26.35d, 23.8d, and 11.0d based on $P_{\rm rot, Noyes}$ and the LC ACFs, the $\vsini$-calculated \emph{$P_{\rm rot,max}$}, and the BIS periodogram, respectively. For K2-222 we ran fits with Gaussian priors on \emph{$P_{\rm{quasi}}$} centered around 28.2d, 17.5d, and 8.0d, based on the polynomial-reduced LC ACF, the $\vsini$-calculated $P_{\rm rot,max}$, and $P_{\rm rot, Noyes}$ plus the RV and FWHM periodogram peaks, respectively. Final exoplanet parameters returned by successful fits agreed within 1-$\sigma$ no matter which priors were used on the GP hyper-parameters. All fits including a Gaussian prior on \emph{$P_{\rm{quasi}}$} just returned the input prior in their posterior, so we report values from the fits including 10-40d and 2-10d Jeffreys priors for K2-79 and K2-222, respectively. Table \ref{radvel_priors} summarizes the priors placed on GP hyper-parameters in our reported {\tt RadVel} fits.

The majority of the {\tt RadVel} fits failed to converge on a solution for the stellar activity signal. RVs for both K2-79 and K2-222 were not optimally scheduled to characterize stellar activity signals, with never more than a single observation taken per night and several multiple-day gaps in each of the analyzed seasons. It is therefore unsurprising that our models have a difficult time characterizing the stellar activity signal. We define \emph{successful} fits as those that converge for values of \emph{A} and \emph{$P_{\rm{quasi}}$}, at least. Even in \emph{successful} fits, posterior distributions show \emph{$\tau$} pushing toward the smallest possible value, consistent with a highly unstable \emph{$P_{\rm{quasi}}$} signal. This was the case for all fits including a GP-generated model with a quasi-periodic kernel. We include a corner plot for each target in the Appendix for those wishing to view the posteriors from the reported {\tt Radvel} fits in further detail (Figures \ref{2237_corner} and \ref{9978_corner}). We also tested the performance of GP fits with a simpler squared-exponential kernel, but found they show the same behavior in the \emph{$\tau$} posteriors as the fits with quasi-periodic kernels and are even less likely to converge on a value for stellar amplitude, \emph{A}.

Table \ref{radvel_post} shows final parameter values returned by \emph{successful} {\tt RadVel} GP fits. The $\approx$24d \emph{$P_{\rm{quasi}}$} signal returned for K2-79 is consistent with that returned by the LC ACFs and close in value to the largest peaks in the LC periodogram, $P_{\rm{rot, Noyes}}$, and the $P_{\rm{rot,max}}$ calculated from $\vsini$. A second smaller peak that shows up in the posterior at 12-13d and the 11d peak in the BIS periodogram both sit near the first harmonic of these signals, as well. We therefore estimate that K2-79 has a $P_{\rm rot}$ in the range 21-26 days. The $\approx$4d \emph{$P_{\rm{quasi}}$} signal returned for K2-222 is consistent with the first harmonics of the estimated $P_{\rm{rot, Noyes}}$ and the $\approx$8d peaks in the RV and FWHM periodograms. While we believe the most likely rotation period of this star is $\approx$8d, estimates of $P_{\rm rot}$ for K2-222 have varied greatly from method to method, and we are less confident in the estimate than that for K2-79. The reported jitter values are low relative to the average RV error for both targets, giving us concern that the GP models could be over-fitting noise that is unrelated to stellar activity. However, the exoplanet semi-amplitudes returned by the \emph{successful} GP fits all have 1-sigma agreement with those in their corresponding {\tt EXOFASTv2} fits. We therefore list the {\tt RadVel} results as further evidence in support of the measured K2-79b and K2-222b semi-amplitudes, but adopt final results for the exoplanet characteristics from our {\tt EXOFASTv2} fits to the full RV data sets.

\begin{table*} 
   \centering
    \caption{Prior probability distributions applied to {\tt Radvel} MCMC fits to K2-79 and K2-222 RVs. The two numbers in each prior are the mean and standard deviation, respectively, for Gaussian prior distributions, or the lower and upper bounds, respectively, for uniform Jeffreys prior distributions. Upper bounds on the eccentricity priors were taken from the posterior distributions returned by {\tt EXOFAST v2} for each respective fit. The Gaussian prior on $\sigma_{\rm HARPS-N}$ is set equal to the distribution of RV errors ($\langle RVerr \rangle$, \hspace{0.05in} $\sigma_{RVerr}$).}\label{radvel_priors}
    \begin{tabular}{lcr} 
        \hline
        Parameter & K2-79 & K2-222 \\
        \hline
        \textbf{Exoplanet} & & \\
        Orbital Period, $P_{\text{b}}$ [days]& Gaussian (10.99470, 0.00045) & Gaussian (15.38857, 0.00088) \\
        Central Transit Time, $T_{\text{C,b}}$ [BJD - 2.4x$10^{6}$] & Gaussian (57103.22750, 0.00084) & Gaussian (57399.0596, 0.0017) \\
        RV Semi-Amp., $K_{\text{b}}$ [m/s] &  Jeffreys (0.5, 10) & Jeffreys (0.5, 10) \\\\
        &&\\
        Eccentricity, $e_{\rm b}$ & & \\
        \hspace{0.1in} All RVs (1-plan) & Jeffreys (0, 0.17) & Jeffreys (0, 0.298)\\
        \hspace{0.1in} S1 RVs (1-plan)  & Jeffreys (0, 0.273) & Jeffreys (0, 0.43)\\
        \hspace{0.1in} S2 RVs (1-plan)  & Jeffreys (0, 0.239) & Jeffreys (0, 0.31)\\
        \hspace{0.1in} All RVs (2-plan) & - & Jeffreys (0, 0.34)\\
        \hspace{0.1in} S1 RVs (2-plan)  & - & Jeffreys (0, 0.38)\\
        \hspace{0.1in} S2 RVs (2-plan)  & - & Jeffreys (0, 0.241)\\
         && \\
        \textbf{Stellar/Noise} & & \\
        HARPS-N jitter, $\sigma_{\rm HARPS-N}$ & Gaussian(4.97, 3.64) & Gaussian(1.80, 0.69) \\
        Stellar Amp., $A$ [m/s] & Jeffreys (0.5, 10) & Jeffreys (0.5, 10) \\
        Evolution Timescale, $\tau$ [days] & Jeffreys (2.0, 100.0) & Jeffreys(2.0, 100.0)\\
        Stellar Period, $P_{\rm quasi}$ [days]  & Jeffreys (10.0, 40.0) & Jeffreys(2.0, 10.0)\\
        High Freq. Variation, $\eta$ & Gaussian (0.25 $\pm$ 0.025) & Gaussian (0.25 $\pm$ 0.025) \\
        \hline
    \end{tabular}
    
\end{table*}


\begin{table*} 
   \centering
    \caption{{\tt Radvel} MCMC posterior probability distributions for exoplanet parameters and GP hyper-parameters fit to the K2-79 and K2-222 RVs.}\label{radvel_post}
    \begin{tabular}{lccr} 
        \hline
         & K2-79 & K2-222 & K2-222 \\
         & All & 1p/All & 1p/S1  \\
        \hline
        \textbf{Exoplanet} & & & \\
        $P_{\text{b}}$ [days]& 10.99475 $\pm$ 0.00047 & 15.3885 $\pm$ 0.00088 & 15.3885 $\pm$ 0.0018 \\
        $T_{\text{C,b}}$ [BJD - 2.4E5] & 57103.2276 $\pm$ 0.0017 & 57399.0594 $\pm$ 0.0032 & 57399.0593 $\pm$ 0.0033 \\
        $K_{\text{b}}$ [m/s] & 3.4 $\pm$ 1.1 & 1.94$^{+0.51}_{-0.53}$ & 1.81 $\pm$ 0.68 \\
        $e_{b}$ & 0.078$^{+0.059}_{-0.054}$ & 0.159$^{+0.093}_{-0.1}$ & 0.21 $\pm$ 0.14  \\
        Arg. of Periastron, $\omega_{\rm b}$ [deg]& -2.7$^{+1.6}_{-2.5}$ & 2.77$^{+1.2}_{-0.99}$ & 2.8 $\pm$ 1.4 \\
         &&&\\
        \textbf{Instrument} & &&\\
        $\gamma_{\rm HARPS-N}$ & -0.22 $\pm$ 0.93 & 0.17$^{+0.41}_{-0.40}$ & 0.35 $\pm$ 0.5 \\
        $\sigma_{\rm HARPS-N}$ & 1.7$^{+1.3}_{-1.6}$ & 1.21$^{+0.52}_{-0.56}$ & 1.39 $\pm$ 0.54 \\
        &&&\\
        \textbf{Stellar Activity} & & &\\
        $A$ [m/s] & 3.1$^{+1.2}_{-1.4}$ & 1.55$^{+0.49}_{-0.56}$  & 1.5$^{+0.60}_{-0.62}$ \\
        $\tau$ [days] & 7.6$^{+11.0}_{-4.3}$ & 13.7$^{+11.0}_{-8.7}$ & 8.7$^{+13.0}_{-5.5}$ \\
        $P_{\rm quasi}$ [days]  & 24.2$^{+9.8}_{9.5}$ & 4.0$^{+3.4}_{-1.2}$ & 4.4$^{+3.7}_{-1.8}$\\
        $\eta$ & 0.252 $\pm$ 0.025 & 0.254 $\pm$ 0.025 & 0.251 $\pm$ 0.025 \\
        \hline
    \end{tabular}
    
\end{table*}

\section{Results and Discussion}\label{discussion}

\begin{table*}
  \caption{Final Derived Exoplanet Characteristics. $\zeta$ is physically related to the chemical composition of an exoplanet core (see Section \ref{discussion} and Figure \ref{MRplot}).}
         \label{plan_chars}
         \small
         \centering
   \begin{tabular}{l | l | l | l }
            \hline
            \noalign{\smallskip}
            Parameter  & K2-79b & K2-222b (1-plan) & K2-222b (2-plan)  \\
            \noalign{\smallskip}
            \hline
            \noalign{\smallskip}
            $P_{\rm orb}$ [days] & 10.99470$^{+0.00031}_{-0.00047}$ & 15.38857 $\pm$ 0.00088 & 15.38857 $\pm$ 0.00088  \\
            $M_{\rm plan}$ [$\mearth$] & 11.8 $\pm$ 3.6 & 8.0 $\pm$ 1.8 & 6.0 $\pm$ 1.9 \\ 
            $R_{\rm plan}$ [$\rearth$] & 4.09$^{+0.17}_{-0.12}$ & 2.35$^{+0.08}_{-0.07}$ & 2.35$^{+0.08}_{-0.07}$  \\
            $\rho_{\rm plan}$ [$\rho_{\Earth}$] & 0.17 $\pm$ 0.06 & 0.62 $\pm$ 0.15 & 0.46 $\pm$ 0.15  \\
            Surface grav. [{\tt g$_{\Earth}$}] & 0.71 $\pm$ 0.22 & 1.45 $\pm$ 0.34 & 1.09 $\pm$ 0.35\\  
            Irradiation [$Flux_{\Earth}$] & 180 $\pm$ 8 & 95 $\pm$ 5 & 95 $\pm$ 5 \\ 
            T$_{eq}$ [K] & 1021$^{+21}_{-20}$ & 878$^{+17}_{-15}$ & 878$^{+17}_{-16}$ \\ 
            $\zeta$ & 2.21 $\pm$ 0.34 & 1.40 $\pm$ 0.16 & 1.50 $\pm$ 0.24 \\
            
            \noalign{\smallskip}
            \hline
     \end{tabular}  
\end{table*}

Table \ref{plan_chars} summarizes the fitted and derived parameters of K2-79b, as well as K2-222b in both the 1-planet and 2-planet cases. Figure \ref{MRplot} shows their location in mass-radius space along with the population of confirmed exoplanets with T$_{eq}$ between 300K and 3000K, 1-sigma mass errors less than 50\%, 1-sigma radius errors less than 20\%, and orbiting stars with radii between 0.6$\rsun$ and 2.0$\rsun$. With radii of 4.09 and 2.35 $R_{\Earth}$, K2-79b and K2-222b exist at two important locations: The \textit{sub-Saturnian desert} and at the upper edge of the \textit{radius valley}, respectively.

 Zeng et al. (2021, accepted) provides evidence for two new dimensionless parameters that can be used to describe exoplanet compositions. The first, $\zeta \equiv (R_{p}/\rearth)(M_{p}/\mearth)^{-1/4}$, can distinguish between three small exoplanet populations: rocky planets composed of varying ratios of silicates and metals ($\zeta \approx$ 1), water worlds dominated by significant amounts of water and  ice ($\zeta \approx$ 1.4), and Neptune/Uranus-like planets with ice-dominated cores hosting small gaseous envelopes ($\zeta \approx$ 2.2) (see histogram in Figure \ref{MRplot}). $\zeta$ is physically related to the chemical composition of the exoplanet core as illustrated by the ternary plot in Figure \ref{MRplot}). The second parameter, $z \equiv \left. \int_{\in \text{envelope}} \frac{dP}{\rho} \right/ \bigg( \frac{G \cdot M_{\oplus}}{R_{\oplus}} \bigg)$, characterizes both the amount and the temperature of the added envelope. We will refer to $\zeta$ and $z$ in the following two sub-sections when discussing potential compositions of K2-79b and K2-222b.  


\subsection{K2-79b}
For K2-79b, we measure a mass and radius of $M_{\rm b}$ = 11.8 $\pm$ 3.6 $\mearth$ and $R_{\rm b}$ = 4.09 $\pm$ 0.17 $\rearth$, which when combined yield a bulk density of $\rho_{\rm b}$ = 0.17 $\pm$ 0.06 $\rho_{\Earth}$. K2-79b falls in the sub-Saturnian desert, where potential compositions are degenerate. If the planet formed water-poor, its predicted composition is a rocky core with an approximately 10\% H/He envelope, by mass \citep{lopez18}. On the other hand, if the planet formed water-rich, the core is likely a rocky and icy combination surrounded by an envelope of mostly H/He gas, and potentially H$_{2}$O vapor \citep{zeng19}. K2-79b receives a relatively high average irradiation level of 180 $\pm$ 8 $F_{\Earth}$ (approximated for circular orbit) and has a low surface gravity of 0.71 $\pm$ 0.22 {\tt g$_{\Earth}$}. Due to the low molecular weight and escape velocity of H$_{2}$ and He, it would be difficult for this exoplanet to maintain a significant water-poor envelope. According to simulations by \citet{lopez18}, K2-79b would not retain a purely H/He envelope over its greater-than-5Gyr lifetime, supporting the presence of heavier gases . 

With a $\zeta$ value of 2.21, K2-79b is likely a Uranus-analog. A slightly H$_{2}$O-enriched atmosphere could explain K2-79b's ability to maintain its envelope while receiving such a high level of irradiation. Assuming the exoplanet hosts a non-cloudy envelope, a pure hydrogen composition would produce a calculated change in transit depth of $\Delta$d = 236.8 parts-per-million (ppm) over five atmospheric scale-heights in transit spectroscopy signals. Assuming a water-enriched envelope made up of 10\% H$_{2}$O and 90\% H$_{2}$, by mole, the transit depth would change 133 ppm over five scale-heights.   


\subsection{K2-222b}
\begin{figure*}[]
\epsscale{1.2}
\plotone{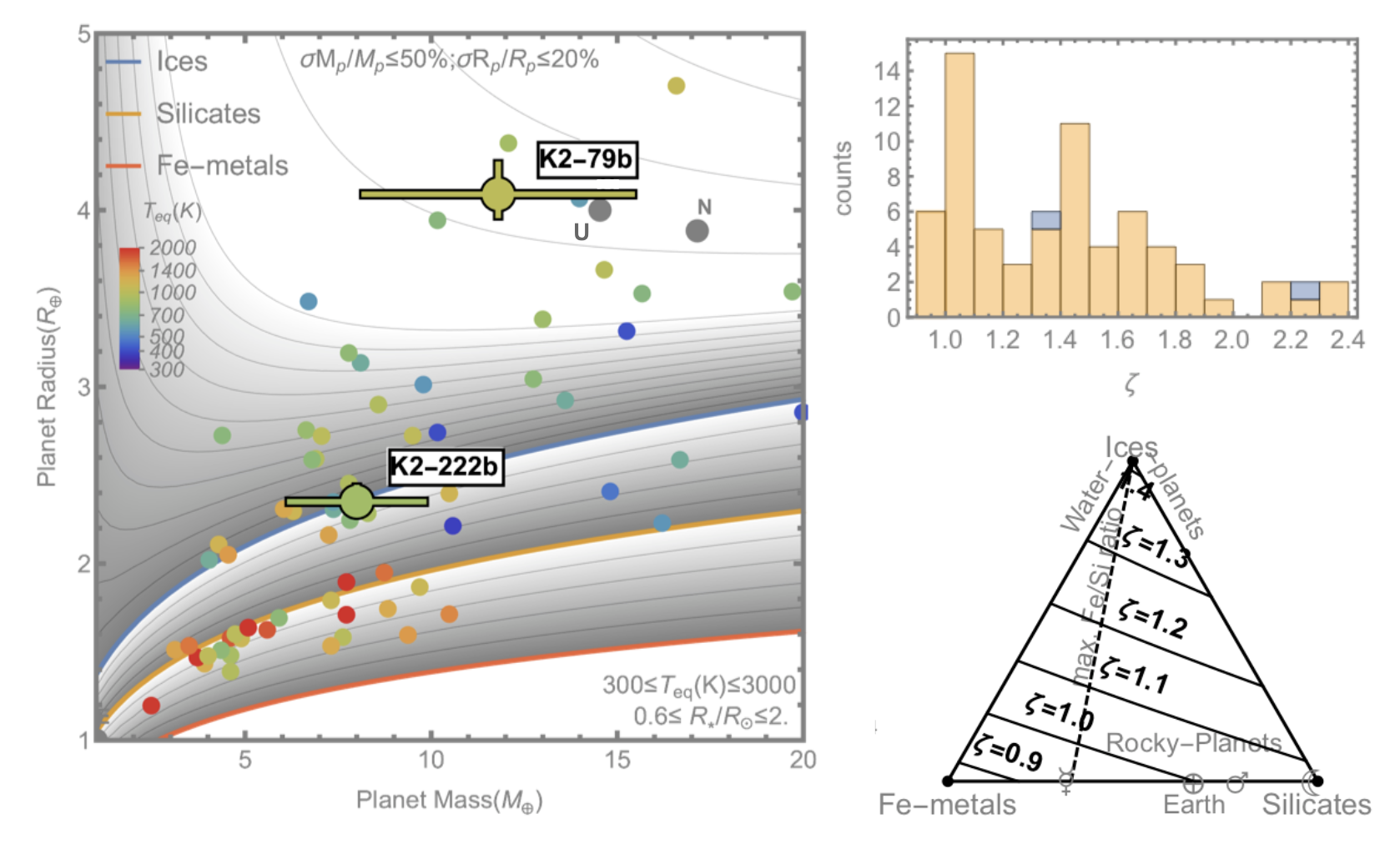}
\caption{The scatter plot (left) shows the placement of K2-79b and K2-222b in mass-radius space among other confirmed small exoplanets (colored points) with T$_{eq}$ between 300K and 3000K, 1-sigma mass errors less than 50\%, 1-sigma radius errors less than 20\%, and orbiting stars with radii between 0.6$\rsun$ and 2.0$\rsun$ (NASA Exoplanet Archive). Uranus (U) and Neptune (N) are indicated by grey dots. The red curve represents Fe-metal compositions of 100\% iron, the orange curve represents pure silicates (MgSiO$_{3}$) composition, and the blue curve represents pure H$_{2}$O compositions. The contours above the blue curve depict different values of the $z$ parameter discussed in Section \ref{discussion}, which characterizes both the amount and the temperature of the added envelope (Section \ref{discussion}; Zeng et al., 2021 (accepted)). The two sets of contours between the three colored curves represent equally spaced mass-proportions of the two compositions they lie between. The histogram (top right) shows the distribution of $\zeta$ values, also discussed in Section \ref{discussion}, for the same population of planets in the scatter plot, with K2-79b and K2-222b shown in blue. The histogram shows clearly a bi-modal distribution, with the first peak ($\zeta \sim 1$) corresponding to purely rocky compositions, while the second peak ($\zeta \sim 1.4$) is consistent with ice-dominated compositions. The histogram also shows a potential tri-modal distribution, with the third peak ($\zeta \sim 2.1-2.4$) consistent with Neptune-like planets. K2-222b is consistent with the second peak, and K2-79b is consistent with the third peak. The ternary plot (bottom left) shows how $\zeta$ relates to various potential exoplanet compositions (Zeng et al., 2021 (accepted)). The solid diagonal lines are contours of $\zeta$, which range from slightly less than 0.9 in one extreme (pure Fe-metals) to slightly more than 1.4 in the other extreme (pure-Ices). The dashed straight line is a fixed ratio of Fe-metals/Silicates, which is a theoretical prediction from maximum collision stripping models.}\label{MRplot}
\end{figure*}

As described in Section 5, there appears to be a second signal in the RVs of K2-222 with a period of 147.5d. AIC and BIC values provide tentative support in favor of that signal being Doppler-induced (Tables \ref{exofit_2237_tab} and \ref{exofit_9978_tab}), but we adopt the 1-planet solution for now, considering the significant phase gaps in the coverage of the 147.5d period returned by our fit. Assuming that the 147.5 signal is Doppler-induced, it would correspond to an object with minimum mass M$\sini$ = 19.3 $\pm$ 4.2 $\mearth$. Including that second signal in our RV fits yields a mass for K2-222b of $M_{\rm b}$ = 6.0 $\pm$ 1.9 $\mearth$. In the case of the adopted single-planet model, we measure a final mass and radius of $M_{\rm b}$ = 8.0 $\pm$ 1.8 $\mearth$ and $R_{\rm b}$ = 2.35 $\pm$ 0.08 $\rearth$, respectively, for K2-222b. Combined, this mass and radius yield a final bulk density of $\rho_{\rm b}$ = 0.62 $\pm$ 0.15 $\rho_{\Earth}$. K2-222b sits just above the upper edge of the radius valley where potential compositions are again degenerate. If it formed water poor, its predicted composition is a rocky core and an approximately 1\% H/He envelope, by mass \citep{lopez18}. If the exoplanet instead formed water-rich, the core is likely equal parts rock and ice hosting an envelope of H$_{2}$O equal to half the core mass \citep{zeng19}. K2-222b receives an irradiation of 95 $\pm$ 5 $F_{\Earth}$ and has a surface gravity of 1.45 $\pm$ 0.34 {\tt g$_{\earth}$}. According to 5 Gyr simulations and its estimated minimum age, K2-222b could potentially retain a pure H/He envelope \citep{lopez18}. However, its $\zeta$ value of 1.40 suggests that K2-222b is most likely a water world (Figure \ref{MRplot}). Assuming an envelope composed of pure hydrogen, we estimate a calculated change in transit depth for K2-222b of $\Delta$d = 158.4 ppm over five scale-heights in transit spectroscopy signals.


\subsection{Utilizing seasonal RV analyses}
With our seasonal RV analyses we were able to measure the masses of K2-79b and K2-222b, in spite of the potentially challenging interference from stellar activity near $P_{\rm orb}$ in both cases. By analysing RV and activity indicator periodograms separately for each available season of data, we determined the stellar activity signal to be evolving from season to season, and identified S1 RVs as suffering from less interference at $P_{\rm rot}$ than S2 RVs. Fitting the full and seasonal RVs separately, we find that all obtained planet parameters agree within errors, providing confirmation that our final mass estimates are not being hindered by stellar activity interference in S2 RVs.

Many RV surveys are now optimizing observations and data reduction approaches to mitigate the effect of stellar activity in the measurement of planet-induced RVs \citep[e.g.][]{LMH, haywood2020, Miklos20, zoe2020, ACC2020}. However, the problem of overlapping exoplanet and stellar activity signals remains difficult to solve in systems where the orbital period of the exoplanet overlaps with a periodic signal from stellar rotation. We suspect the key observing strategy with such targets will be to observe with high-cadence (at least once per night) over at least two separate seasons, knowing that the planet signal does not change, but the average spot distribution, and therefore stellar activity signal, does. In cases of highly stable activity signals, several seasons of separation between observations may be favorable in order to observe different average spot distributions. We hope that future simulations will confirm or rule-out these potential strategies. However, much of the archival RV data was collected without the inclusion of optimal strategies for mitigating stellar activity signals. In those cases, an analysis approach like the one described in this paper might help to extract exoplanet signals.


\clearpage

\section*{Acknowledgements}
The HARPS-N project was funded by the Prodex Program of the Swiss Space Office (SSO), the Harvard- University Origin of Life Initiative (HUOLI), the Scottish Universities Physics Alliance (SUPA), the University of Geneva, the Smithsonian Astrophysical Observatory (SAO), the Italian National Astrophysical Institute (INAF), University of St. Andrews, Queen’s University Belfast, and University of Edinburgh. This work have been supported by the National Aeronautics and Space Administration under grant No. NNX17AB59G, issued through the Exoplanets Research Program. Parts of this work have been supported by the Brinson Foundation. AMo acknowledges support from the senior Kavli Institute Fellowships. CAW acknowledges support from Science and Technology Facilities Council (STFC) grant ST/P000312/1. S.H.S. acknowledges support by NASA Heliophysics LWS grant NNX16AB79G. L.Z. acknowledges support by the Sandia Z Fundamental Science Program by the Department of Energy National Nuclear Security Administration under Awards DE-NA0003904 (to S.B.J.) (principal investigator) with Harvard University. 

This work has made use of data from the European Space Agency (ESA) mission {\it Gaia} (\url{https://www.cosmos.esa.int/gaia}), processed by the {\it Gaia}Data Processing and Analysis Consortium (DPAC,
\url{https://www.cosmos.esa.int/web/gaia/dpac/consortium}). Funding for the DPAC has been provided by national institutions, in particular the institutions participating in the {\it Gaia} Multilateral Agreement.

We also thank Sarah Blunt for graciously answering questions regarding use of the {\tt RadVel} software. 

\clearpage

\appendix

\begin{figure*}[h]

\centering
\includegraphics[width=1.0\textwidth]{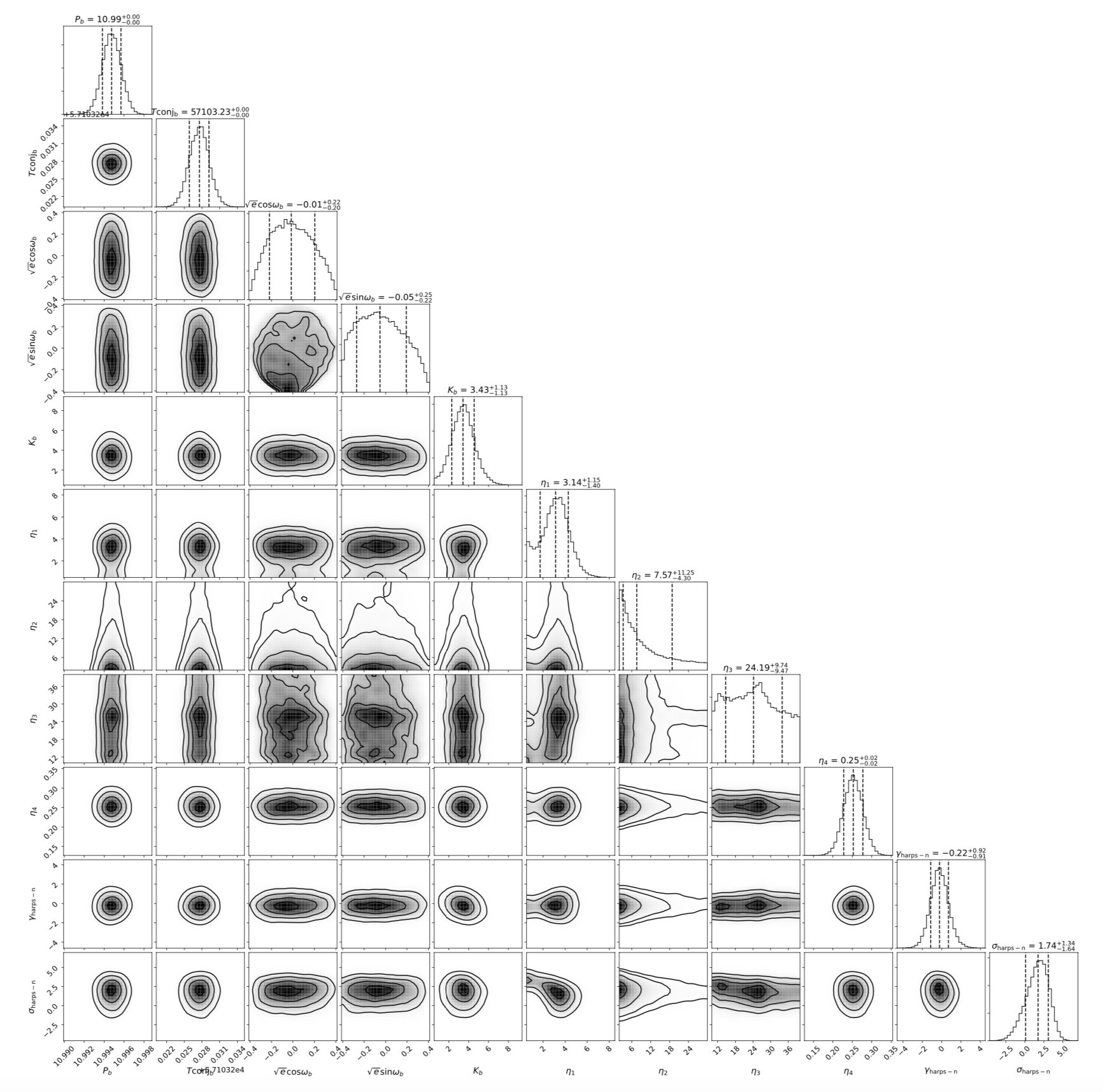}
\caption{Corner plot showing posterior distributions of hyper-parameters from the \emph{successful} K2-79 {\tt Radvel} MCMC fits including a stellar activity signal modelled with quasi-periodic Gaussian Processes regression. The {\tt Radvel} hyper-parameters 1-4 represent \emph{A}, \emph{$\tau$}, \emph{$P_{\rm{quasi}}$}, and \emph{$\eta$}. The $\approx$24d \emph{$P_{\rm{quasi}}$} signal is consistent with that returned by the LC ACFs, as well as close in value to the largest peaks in the LC periodogram, $P_{\rm rot, Noyes}$, and the $P_{\rm rot,max}$ calculated from $\vsini$. The second smaller peak that shows up in the posterior at 12-13d and the 11d peak in the BIS periodogram both sit near the first harmonic of the 24d signal. We therefore estimate that K2-79 has a $P_{\rm rot}$ in the range 21-26 days.}
\label{2237_corner}
\end{figure*}

\begin{figure*}[h]
  \centering
  \includegraphics[width=1.0\textwidth]{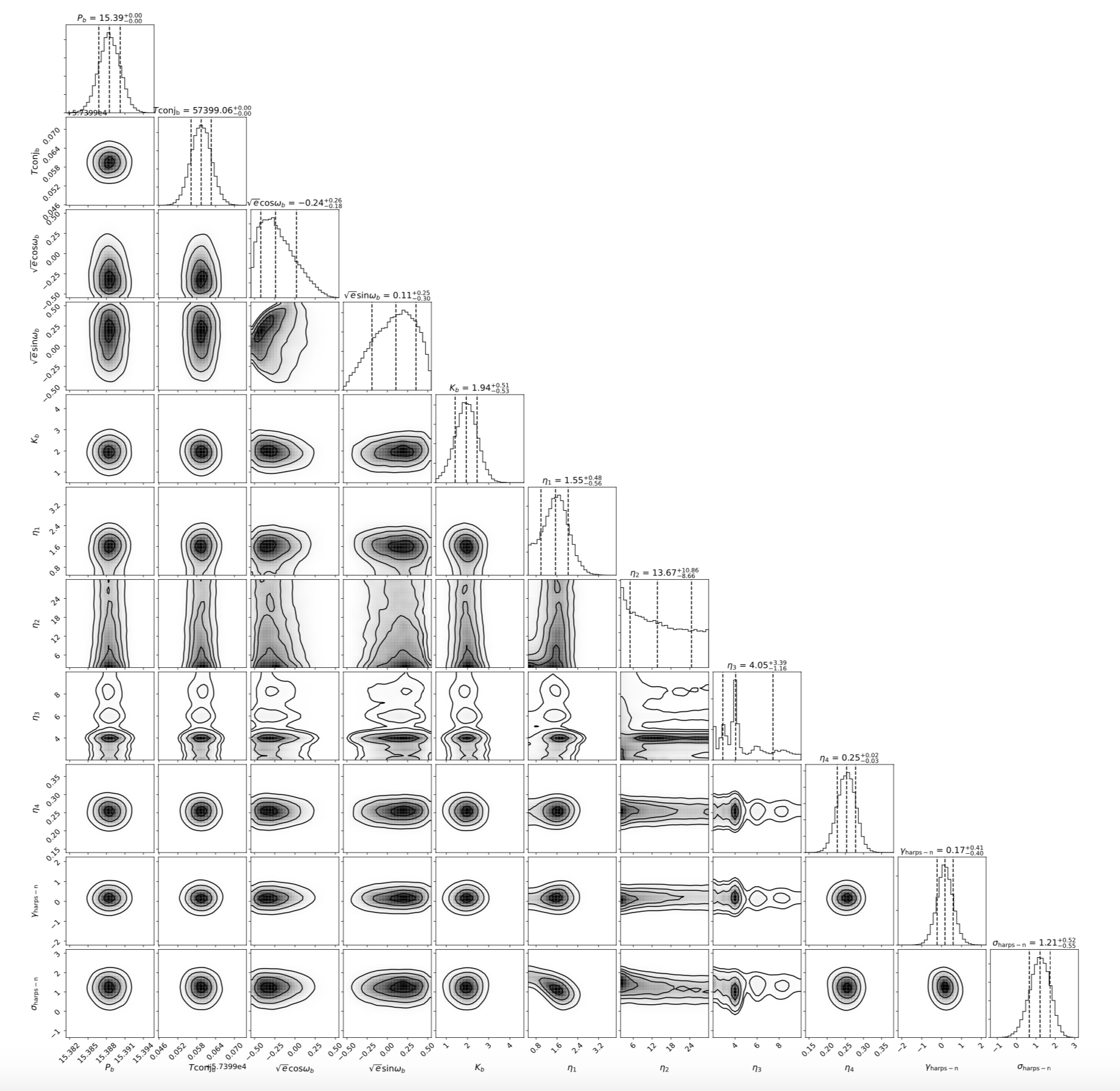}
\caption{Corner plot showing posterior distributions of hyper-parameters from one of the \emph{successful} K2-222 {\tt Radvel} MCMC fits including a stellar activity signal modelled with quasi-periodic Gaussian Processes regression. The {\tt Radvel} hyper-parameters 1-4 represent \emph{A}, \emph{$\tau$}, \emph{$P_{\rm{quasi}}$}, and \emph{$\eta$}. We only show the corner plot for the fit utilizing all RVs because that from the fit to only S1 RVs contains no notable differences. The $\approx$4d \emph{$P_{\rm{quasi}}$} signal returned for K2-222 is consistent with the first harmonics of the estimated $P_{\rm rot, Noyes}$ and the $\approx$8d peaks in the RV and FWHM periodograms. While we believe $\approx$8d is the most likely rotation period of this star, estimates of $P_{\rm rot}$ for K2-222 have varied greatly for certain methods, and we are less confident in this estimate than that for K2-79.}
\label{9978_corner}
\end{figure*}

\newpage
\bibliography{Nava_etal_2021.bib}


\clearpage



\clearpage






\end{document}